\documentclass{jfm}

\usepackage{graphicx}
\usepackage{newtxtext}
\usepackage{newtxmath}
\usepackage{natbib}
\usepackage{hyperref}
\hypersetup{
    colorlinks = true,
    urlcolor   = blue,
    citecolor  = black,
}

\newcommand{\RomanNumeralCaps}[1]

\title{Towards realistic ocean flow decomposition using optimal balance with time-averaging}

\author{Silvano G. Rosenau\aff{1} 
  \corresp{\email{silvano.rosenau@uni-hamburg.de}},
  M. Chouksey\aff{1,2}
  \and C. Eden\aff{1}}

  \affiliation{\aff{1}Institut f\"ur Meereskunde, Universit\"at Hamburg, Germany
\aff{2}Leibniz Institut f\"ur Ostseeforschung Warnem\"unde, Germany}

\begin{document}
\maketitle


\begin{abstract}
Decomposing oceanic and atmospheric flow fields into their slowly evolving balanced components and fast evolving wave components is essential for understanding processes like spontaneous wave emission. 
To study these processes, the decomposition into the linear geostrophic (slow) and non-geostrophic (fast) components is not precise enough.
More precise methods, such as optimal balance and nonlinear normal mode decomposition, account for nonlinear effects, but their application has so far been limited to idealized model configurations that exclude irregular lateral boundaries, depth variations, a varying Coriolis parameter, or other environmental conditions.
Here, we address these limitations by modifying the optimal balance method such that it is applicable to more realistic model setups. 
The modification employs a time-averaging procedure to project onto the linear geostrophic component, eliminating the need for a Fourier transformation as required in the original optimal balance method.
We demonstrate analytically and experimentally that the new method converges to the original method when either the time-averaging period or the number of optimal balance iterations increases. We test the new method using a two-dimensional single-layer model and a three-dimensional non-hydrostatic model with varying initial conditions and Rossby numbers ranging from 0.03 to 0.5. 
In all tested configurations, the differences between balanced states obtained from the new method and those from the original method become exponentially small. Optimal balance with time-averaging thus comes forth as a promising new flow decomposition tool for complex flows.
\end{abstract}

\begin{keywords}
Authors should not enter keywords on the manuscript, as these must be chosen by the author during the online submission process and will then be added during the typesetting process (see \href{https://www.cambridge.org/core/journals/journal-of-fluid-mechanics/information/list-of-keywords}{Keyword PDF} for the full list).  Other classifications will be added at the same time.
\end{keywords}



\section{Introduction}
\label{sec:intro}

Geophysical flows can be understood as a two timescale system consisting of a slow and a fast component. The need to decompose a flow into its slowly evolving balanced component and its fast wave components first emerged in geophysical research as a way to minimize artificial, strong, fast oscillations caused by gravity waves in weather forecast models.
These fast oscillations occur when initializing numerical models with observations \citep{temperton1981normal} and degrade the quality of weather forecasts.
Although that problem was resolved by other means than explicit flow decomposition \citep{lynch1992initialization}, there has recently been a resurgence of interest in flow decomposition methods, particularly concerning the determination of energy transfer between different scales and dynamical regimes like waves and geostrophically balanced flow.
Energy is transferred from the atmosphere to the ocean predominantly by wind forcing. 
Once in the ocean, this energy tends to propagate towards larger scales within the inverse energy cascade \citep{Scott:05}. 
To maintain an equilibrium state, energy must be removed from the large scales, which are typically in geostrophic balance and evolve slowly. 
Several processes have been proposed for this, one of which is the spontaneous wave emission from the balanced flow (e.g. \citealp[]{Vanneste:13,Chouksey:18}).  This loss of geostrophic balance on large scales would lead to an energy transfer to gravity waves, then within the wave field to a downscale energy transfer, and finally collapse to small-scale turbulence and molecular dissipation (e.g. \citealp[]{molemaker2005}).
In the atmosphere, the process of spontaneous wave emission is thought to be a significant source of gravity waves \citep{hertzog2008estimation,plougonven2014internal}.
If dissipation by spontaneous wave emission plays a relevant role in the oceanic or atmospheric energy budgets, this process needs to be parameterized in general circulation models, as it might potentially be unresolved. 
However, before developing parameterizations, we first need to understand under which conditions waves are generated and how much energy is transferred.
Answering these questions requires a precise flow decomposition method, which is the focus of this study.

Several decomposition methods have been proposed, either by addressing nonlinearity directly, such as the nonlinear normal mode decomposition (NNMD) \citep{Machenhauer:77,WarBokShe95,Eden:2019gravity} or the optimal balance (OB) \citep{masur2020optimal} modified from the optimal potential vorticity balance \citep{Viudez:04}, or by simple temporal filtering, such as the digital filter initialization \citep{lynch1992initialization} or the particle-based Lagrangian filtering \citep{shakespeare2016spontaneous}, which offer less accuracy.
However, no decomposition method currently exists that simultaneously works in realistic model setups (with irregular boundaries, varying depth, and variable Coriolis parameter) and decomposes the flow with accuracy high enough to study spontaneous wave emission, like NNMD or OB.
In this work, we present a modification of the OB method that can be used in models that include changes in depth and irregular lateral boundaries.
With this modification, studying energy transfers between waves and balanced flow becomes possible in more realistic ocean models.

In the following sections, we first introduce the system of equations and explain the concept of the slow (geostrophically balanced) manifold.
Using that concept, in Section \ref{sec:flow_decomposition}, we introduce the OB method and explain how it works.
We then present the novel OBTA \textbf{acronym not introduced before} method and examine its convergence behavior from an analytical perspective.
Thereafter, we validate the convergence behavior in a numerical two-dimensional single-layer model and a three-dimensional non-hydrostatic model. Key conclusions are summarized in the final section.

\section{Theory}
\label{sec:theory}

The concept of OB is based on the idea of a slow manifold.
An intuitive understanding of the slow manifold helps in comprehending how and why these decomposition methods work.
Therefore, we first introduce and discuss the slow manifold before detailing the decomposition methods in Section \ref{sec:flow_decomposition}.
Although, for simplicity, we use a two-dimensional single-layer model (shallow water model) in this and the next section, the concepts are not limited to this model and can be readily extended to the three-dimensional primitive equations or the three-dimensional non-hydrostatic model used later in this work.

\subsection{Shallow water model equations}
\label{subsec:shallow_water_model}

We consider the two-dimensional scaled equations for the flow of a single layer (shallow water) model in the double periodic domain $\Omega = [0,2\upi]\times[0,2\upi]$ on the $f$-plane, given by:
\begin{equation}
  \partial_t \boldsymbol{z} = 
  \mathcal{L} \bcdot \boldsymbol{z} + Ro \, \boldsymbol{N} (\boldsymbol{z}),
  \label{eq:SWeq}
\end{equation}
with:
\begin{equation}
  \boldsymbol{z} = \begin{pmatrix} u \\ v \\ h \end{pmatrix}, \quad
  \mathcal{L} = \begin{pmatrix} 0 & f & -\partial_x \\ 
    -f & 0 & -\partial_y \\
    -c^2 \partial_x & -c^2 \partial_y & 0 
  \end{pmatrix}, \quad
  \boldsymbol{N} (\boldsymbol{z}) = 
  \begin{pmatrix} -(u \partial_x + v \partial_y) u \\ 
    -(u \partial_x + v \partial_y) v \\ 
    -\partial_x (uh) - \partial_y (vh)  
  \end{pmatrix}
  \label{eq2}
\end{equation}

The state vector $\boldsymbol{z}$ contains the horizontal velocity components $u$ and $v$ and the layer thickness $h$.
The linear operator $\mathcal{L}$ contains the constant Coriolis parameter $f$ and the Burger number $c$.
The non-linear term $\boldsymbol{N}(\boldsymbol{z})$ contains the advection terms.
The Rossby number $Ro$ is a measure of the non-linearity of the system and is given by $Ro = U/(fL)$, where $U$ is the typical velocity scale and $L$ is the typical length scale of the system. 
We note that this system is identical to the single-layer model discussed by \citet[their Section 2.1, Eq. 2.3]{chouksey2023comparison}

The dynamics of a state vector $\boldsymbol{z}$ for the linear system with $Ro = 0$ can be described as a superposition of two inertia-gravity wave modes and one geostrophic mode.
The geostrophic mode remains constant in time since $f = \text{const}$, while the inertia-gravity wave modes oscillate rapidly.
For varying $f$, the geostrophic mode evolves slowly and becomes the Rossby wave mode.
The decomposition of an arbitrary state into its geostrophic and non-geostrophic components is a key step in the OB decomposition method.
To achieve this, we first apply a Fourier transformation to the shallow water equation, which yields:
\begin{equation}
  \partial_t \hat{\boldsymbol{z}} = 
  -\mathrm{i} \mathsfbi{A} \bcdot \hat{\boldsymbol{z}} 
  + Ro\, \hat{\boldsymbol{N}}(\boldsymbol{z})
  , \quad
  \hat{\boldsymbol{z}} = \mathcal{F}(\boldsymbol{z})
  , \quad
  \hat{\boldsymbol{N}} = \mathcal{F}(\boldsymbol{N})
  \label{eq:spec_end}
\end{equation}
with:
\begin{equation}
  \mathcal{F}(\boldsymbol{z}) =
  \frac{1}{4\upi^2} \int_0^{2\upi} \boldsymbol{z} e^{-\mathrm{i}\boldsymbol{k}\bcdot\boldsymbol{x}} \,d\boldsymbol{x}
  , \quad
  \mathsfbi{A} = 
  \begin{pmatrix} 
    0 & \mathrm{i} f & k_x \\ 
    -\mathrm{i}f & 0 & k_y \\
    c^2 k_x & c^2 k_y & 0 
  \end{pmatrix}
  \label{eq:Ak_SW}
\end{equation}
where $\boldsymbol{k} = (k_x, k_y)$ denotes the wavenumber vector.

Note that the linear matrix $\mathsfbi{A}$ and its eigenvalues and eigenvectors change when replacing the continuous differential operators with the finite difference operators that we use in the numerical model, see Section \ref{sec:numerical_experiments}.
In the numerical experiments detailed below, we use the discrete eigenvalues and eigenvectors.
However, for the sake of simplicity, we only consider the continuous case in this section.
For more details on the discrete case, we refer the reader to \citet[Section 3.7]{Chouksey:18b}.
The eigenvalues for the continuous case are given by:
\begin{equation}
  \omega_0 = 0
  ~~, \hspace{2cm} 
  \omega_{1,2} = \pm \sqrt{f^2 + c^2 (k_x^2 + k_y^2)}
\end{equation}
The eigenvalue $\omega_0$ corresponds to the frequency of the linear geostrophic mode.
This frequency is zero here, since we set $f = \text{const}$.
When considering a varying $f$, $\omega_0$ becomes the dispersion relation of Rossby waves, see Appendix B of \citet{eden2019mixed}.
The other two eigenvalues, $\omega_{1,2}$, correspond to the dispersion relation of linear inertia-gravity waves.

To decompose an arbitrary state $\hat{\boldsymbol{z}}$ into the linear modes, we need to project it onto the corresponding eigenspaces $\mathbb{E}_i$:
\begin{equation}
  \mathbb{E}_i = 
  \left\{
    \left.
      \hat{\boldsymbol{z}} \in \mathbb{C}^3 ~
    \right| 
    ~\mathsfbi{A} \bcdot \hat{\boldsymbol{z}} = 
    \omega_i \hat{\boldsymbol{z}}
  \right\} 
  = \text{span}(\boldsymbol{q}_i),
  \label{eq:eigenspaces}
\end{equation}
where $\boldsymbol{q}_i$ are the eigenvectors of $\mathsfbi{A}$, given by Eq.~(\ref{eq:eigenvectors}), and $i = 0, 1, 2$.
Since the eigenvectors $\boldsymbol{q}_i$ are orthogonal to each other, the projection onto the eigenspaces $\mathbb{E}_i$ is given by the projection matrix $\mathsfbi{P}$:
\begin{equation}
  \hat{\boldsymbol{z}}_i = \mathsfbi{P}_i \bcdot \hat{\boldsymbol{z}}
  ~~, \hspace{2cm}
  \mathsfbi{P}_i = \frac{1}{\boldsymbol{p}_i^* \bcdot \boldsymbol{q}_i}
  \boldsymbol{q}_i \bcdot \boldsymbol{p}_i^*,
  \label{eq:spectral_projection}
\end{equation}
(no summation over $i$ implied here) with:
\begin{eqnarray}
  \boldsymbol{q}_i = 
  \begin{pmatrix}
    \omega_i \boldsymbol{k} + \mathrm{i} f \underset{\neg}{\boldsymbol{k}} \\
    f^2 - \omega_i^2
  \end{pmatrix} 
  ~~, \hspace{0.75cm}
  \boldsymbol{p}_i = 
  \begin{pmatrix}
    c^2 \omega_i \boldsymbol{k} - \mathrm{i} f c^2 \underset{\neg}{\boldsymbol{k}} \\
    f^2 - \omega_i^2
  \end{pmatrix}
  ~~, \hspace{0.75cm}
  \underset{\neg}{\boldsymbol{k}} = \begin{pmatrix}
    -k_y \\
    k_x
  \end{pmatrix},
  \label{eq:eigenvectors}
\end{eqnarray}
where $\hat{\boldsymbol{z}}_i$ are the projections on the linear normal modes, the star \textbf{*} denotes the Hermitian transpose, and $\boldsymbol{p}_i$ are the left eigenvectors of $\mathsfbi{A}$. 
The left eigenvectors satisfy:
\begin{equation}
  \boldsymbol{p}_i^* \mathsfbi{A} = \omega_i \boldsymbol{p}_i^*,
  \hspace{1cm} \implies \hspace{1cm}
  \mathsfbi{P}_i \bcdot \mathsfbi{A} = \omega_i \mathsfbi{P}_i.
  \label{eq:projection_of_matrix}
\end{equation}

We use the latter property to obtain the spectral shallow water equation in its linear normal mode representation.
For this, we multiply the projection matrix $\mathsfbi{P}_i$ with the spectral shallow water equation Eq.~(\ref{eq:spec_end}) and insert Eq.~(\ref{eq:projection_of_matrix}), which yields:
\begin{equation}
  \partial_t \hat{\boldsymbol{z}}_i = - \mathrm{i} \omega_i \hat{\boldsymbol{z}}_i
  + Ro \, \hat{\boldsymbol{N}}_i (\boldsymbol{z})
  \hspace{1.0cm} \text{with} \hspace{0.5cm}
  \hat{\boldsymbol{N}}_i (\boldsymbol{z}) = \mathsfbi{P}_i \bcdot \hat{\boldsymbol{N}} (\boldsymbol{z}),
  \label{eq:linear_normal_mode_representation}
\end{equation}
The linear normal modes are decoupled from each other in the linear model with $Ro = 0$, but in the nonlinear model with $Ro > 0$, all modes are coupled.
The question of the existence of a slow manifold corresponds to finding a projection similar to $\mathsfbi{P}_i$ such that the resulting slow geostrophic mode is independent of the others in the full nonlinear model.
The decomposition methods OB, OBTA, and NNMD aim to provide such a projection \cite{chouksey2023comparison}.


\subsection{The slow manifold}
\label{subsec:slow_manifold}

The slow manifold $\mathcal{S} \subset \Gamma$ is a hypothetical subset of the phase space $\Gamma$, where $\Gamma$ contains all possible states $\hat{\boldsymbol{z}}$ (or $\boldsymbol{z}$) in the system of equations.
The term manifold means, loosely speaking, that $\mathcal{S}$ can be parameterized from spaces with a lower dimension than the phase space $\Gamma$.
An example of a manifold is the circle in two dimensions, which can be parameterized from the one-dimensional space of angles.
The term slow refers to the time evolution of the states on the manifold, which should be slow compared to the fastest waves in the system.
In the shallow water equations, the fastest waves are inertia-gravity waves with a frequency lower bounded by the Coriolis frequency $f$.
Hence, the time derivative of all states in $\mathcal{S}$ should be smaller than the Coriolis frequency $f$:
\begin{equation}
  \left|\left|\partial_t \hat{\boldsymbol z} \right|\right| < ||f \hat{\boldsymbol z}|| 
  \hspace{1.5cm} \forall \, \hat{\boldsymbol z} \in \mathcal{S}
  \label{eq:slow_manifold_is_slow}
\end{equation}
Here, $|| \cdot ||$ denotes a general norm, which we do not further specify as its exact form is not crucial for the scope of this discussion.
For more details on the definition of the slow manifold, we refer the reader to \citet{Vanneste:13} and the references therein.

A manifold $\mathcal{S}$ is called invariant when it does not change under the action of the dynamical system, i.e.
\begin{eqnarray}
  \Phi^{t^*} \hat{\boldsymbol z} & \in & \mathcal{S} \hspace{1.5cm} \forall \, \hat{\boldsymbol z} \in \mathcal{S} \,\, \land \,\, \forall \, t^* \in \mathbb{R}^+ 
  \label{eq:slow_manifold_is_invariant}
  \\
  \text{with} \hspace{1.5cm} \Phi^{t^*} \hat{\boldsymbol z} & = & \int_{t_0}^{t_0 + t^*} \partial_t \hat{\boldsymbol{z}} dt
  \label{eq:evolution}
\end{eqnarray}
where $\Phi^{t^*}$ is the evolution operator that takes a state $\hat{\boldsymbol{z}}$ and integrates it to the new time $t_0 + t^*$. In other words, Eq.~\ref{eq:slow_manifold_is_invariant} states that the manifold $\mathcal{S}$ is called invariant when any state $\hat{\boldsymbol z}$ on $\mathcal{S}$ stays on $\mathcal{S}$ when 
$\hat{\boldsymbol z}$ it is integrated forward in time. 
Let us consider the simplest case of the linear system for $Ro = 0$, where the three eigenspaces $\mathbb E_i$ from Eq.~\ref{eq:eigenspaces} are invariant manifolds since
\begin{equation}
  \Phi^{t^*} \hat{\boldsymbol z}_i = \hat{\boldsymbol z}_i e^{-\mathrm{i}\omega_i t^*}
  \in \mathbb E_i \hspace{1.5cm} \forall \, \hat{\boldsymbol z}_i \in \mathbb E_i  \,\, \land \,\, \forall \, t^* \in \mathbb{R}^+
  \label{eq:general_solution}
\end{equation}
For this system, the linear geostrophic eigenspace $\mathbb E_0$ is the slow manifold, since the time derivative of any state in $\mathbb E_0$ is zero.

However, when considering the non-linear system with $Ro>0$, the slow manifold is less trivial to find. 
The linear geostrophic mode is coupled to the linear inertia-gravity wave mode by the nonlinear term. This implies that $\mathbb E_0$ is no longer invariant. Inertia-gravity waves can be generated by nonlinear interactions of the geostrophic mode with itself
\begin{equation}
  \mathsfbi{P}_{1,2} \bcdot \partial_t \hat{\boldsymbol{z}}_0 = Ro \, \hat{\boldsymbol{N}}_{1,2}(\hat{\boldsymbol{z}}_0) \neq 0 \hspace{1.5cm} \hat{\boldsymbol z}_0 \in \mathbb E_0
  \label{eq:wave_from_geo}
\end{equation}
This term, even though it may seem so at first glance, cannot be understood as spontaneous wave emission. This will be further discussed later in this section.

\begin{figure}
  \centerline{\includegraphics[width=\linewidth]{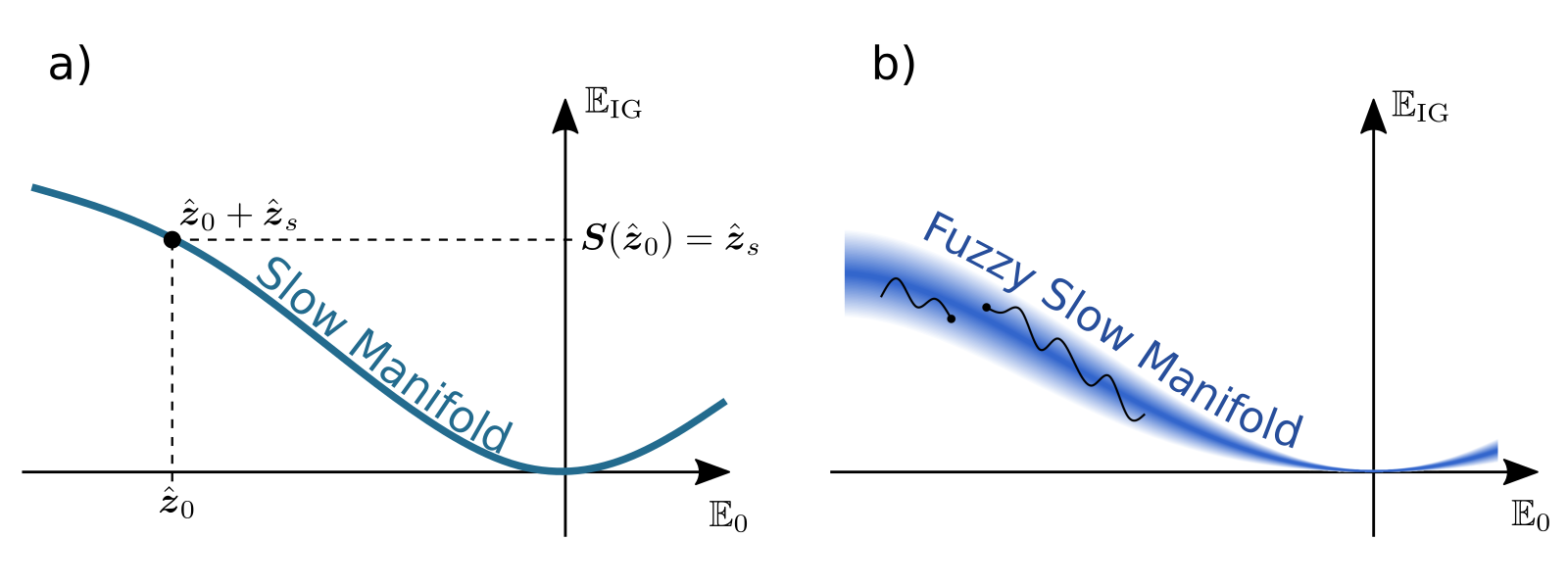}}
  \caption{Illustration of the slow manifold in the phase space. The horizontal axis is the magnitude of the linear geostrophic component and the vertical axis the magnitude of the linear non-geostrophic component.  a) shows a schematic of a slow manifold $\mathcal{S}$ that can be parameterized with the slaved mode projector $\boldsymbol S$. b) illustrates the concept of a fuzzy slow manifold. The black dots are states that slowly evolve around the fuzzy slow manifold.}
  \label{fig:slow_manifold}
\end{figure}

The question regarding the existence of an invariant slow manifold in the nonlinear dynamical system is still open. 
\citet{Leith:80}  proposes a manifold, in which the linear geostrophic mode and the linear inertia-gravity wave mode balance each other, such that their time derivative remains small. 
The linear inertia-gravity wave mode that is in balance with the geostrophic mode is called the {\it slaved mode} $\hat{\boldsymbol z}_s$
\begin{equation}
  \hat{\boldsymbol z}_s = (\mathsfbi{P}_1 + \mathsfbi{P}_2) \bcdot \hat{\boldsymbol z} \hspace{1.5cm} \text{for} \hspace{0.4cm} \hat{\boldsymbol z} \in \mathcal{S}
  \label{eq:SlaveMode}
\end{equation}
Assuming that an invariant slow manifold exist and that for each linear geostrophic mode there is exactly one slaved mode, the slow manifold could be parameterized as a function of the linear geostrophic subspace. We refer to this function as the slaved mode projector $\boldsymbol S$
\begin{equation}
  \boldsymbol S: \mathbb E_0 \rightarrow \mathbb E_\text{IG} \hspace{1.5cm} \text{with} \hspace{0.4cm}
  \hat{\boldsymbol z}_0 + \boldsymbol S(\hat{\boldsymbol z}_0) = \hat{\boldsymbol z}_0 + \hat{\boldsymbol z}_s \in \mathcal{S}
\end{equation}
where $E_\text{IG} = \mathbb E_1 + \mathbb E_2$ denotes the linear inertia-gravity wave subspace.
A schematic of this parameterization is given in Fig.~\ref{fig:slow_manifold}.

A state that lies on the slow manifold will slowly evolve on the slow manifold, while a state that lies away from the slow manifold oscillate quickly around it. Consequently, when starting on the linear geostrophic eigenspace, the state may move towards the slow manifold. This movement is the term that we observe in Eq.~(\ref{eq:wave_from_geo}). 
Hence, this term does not represent spontaneous wave emission but instead it is a result of the geostrophic mode only being 
an approximation of the slow manifold.

Slow manifolds were found for some simple idealized dynamical systems \citep{roberts1985simple}. However, the search of an invariant slow manifold in complex dynamical system, such as realistic ocean models, has been unsuccessful so far. \citet{lorenz1987nonexistence} investigated a simple 5-component system and argued that a slow manifold does not exist for this simple system. It is now generally accepted that, in realistic models, an invariant slow manifold does not exist \citep{lynch,Vanneste:13}, but a proof is missing.
Instead, some are pursuing the idea of a fuzzy slow manifold \citep{Lorenz:86}, also referred to as the slow quasi-manifold \citep{mcintyre1996hamiltonian}.
This is a loosely defined set of states for which the trajectory of each state on the set stays exponentially close to the set. Additionally, states on the fuzzy slow manifold should evolve slowly. We could think of a slow balanced signal that emits a small signal that oscillates quickly around the balanced signal. Fig.~\ref{fig:slow_manifold} illustrates the idea of a fuzzy slow manifold. 
Strictly speaking, the fuzzy slow manifold is neither invariant, nor slow, nor a manifold in a mathematical sense, but could be considered as a quasi manifold. When we refer to the slow manifold in the following work, we actually mean the fuzzy slow manifold in the likely case of the non-existence
of a slow manifold in the mathematical sense.

\section{Flow decomposition}
\label{sec:flow_decomposition}

\subsection{The optimal balance (OB) method by Masur and Oliver (2020)}\label{sec:OB}

The OB concept for flow decomposition was introduced by \citet{Viudez:04} in the specific setting of semi-Lagrangian potential vorticity based numerical models. This method was later generalized for use in other models by  \citet{masur2020optimal}.
The concept of OB is based on the idea that the (fuzzy) slow manifold describes a (more or less) stable equilibrium. If a state is pushed away from that equilibrium, it will accelerate back towards the slow manifold. In OB this principle is reversed: Instead of pushing the state away from the slow manifold, the manifold itself gets deformed such that the state follows the deformation. This deformation is achieved by artificially modifying the amplitude of the nonlinear term in Eq.~(\ref{eq:SWeq}) with the scaling factor $\rho$
\begin{equation}
  \partial_t \boldsymbol z = \mathcal{L} \bcdot \boldsymbol z + \rho Ro\, \boldsymbol N(\boldsymbol z)
\end{equation}
For $\rho = 0$ the slow manifold is known, it is  the linear geostrophic eigenspace. 
Hence, when ramping the nonlinear term to zero during a model integration,
projecting the obtained state on the linear geostrophic eigenspace, and then ramping the nonlinear term back to $\rho = 1$ in a subsequent integration reversed in time, we should end up with a state that is closer to the slow manifold. 
During the ramping, the time evolves so that the state can follow the deformation of the manifold. The duration of the ramping is called ramp period $\tau$ and significantly affects the accuracy of the balancing. A detailed discussion on the choice of the ramp time $\tau$ and function $\rho(t)$  can be found in \citet{masur2020optimal}. In short, 
$\tau$ should neither be too short nor too long, depending on the model configuration.
We performed sensitivity experiments of the ramp period and find good results for a ramp period of $\tau = 5/Ro$ for our configurations. The results of the ramp period are in general agreement with the findings of \citet{masur2020optimal} and are not shown here.

The first time integration is backwards in time and the second time integration is forward in time. Otherwise, when ramping a balanced state forward in time towards the linear end, and then forward in time towards the nonlinear end, it would end up at a different point. 
To ensure the feasibility of backward time integrations, certain requirements must be met by the model. For instance, dissipation and friction must be disabled during the ramping since they inherently operate in a forward time direction only. Additionally, we may encounter problems with 
highly non-linear processes like convection
or parameterisations of it.
A preliminary assessment on whether OB is applicable can be obtained by ramping a state to the linear end and subsequently back to the nonlinear end without any projection onto the geostrophic mode at the linear end. The error of balancing will likely be at least as large as the difference between the initial state and the backward-forward ramped state. 

\noindent The ramp function $\rho(t)$ should be continuous and differentiable. Further, \citet{GottwaldMO:2017:OptimalBA} argued that all derivatives should vanish at both the linear end ($\rho=0$) and at the nonlinear end ($\rho=1$)
\begin{equation}
  \frac{\partial^i \rho(t)}{\partial t^i} = 0 
  ~~~~~~ \text{at} ~~~ t=0, \tau ~~~~~~ \text{for} ~~~ i=1,2,\dots
  \label{eq:vanishing_derivative}
\end{equation}
We use an exponential ramp function $\rho(t)$ which satisfies all the above mentioned requirements,
given by
\begin{equation}
  \rho(t) = \frac{e^{-\tau/t}}{e^{-\tau/t} + e^{-\tau/(\tau - t)}} 
  \label{eq:expo_ramp}
\end{equation}
A comparison of different ramp functions can be found in \citet{masur2020optimal}, where best results were obtained with the exponential function that we use here.

\begin{figure}[t]
  \centerline{\includegraphics[width=0.6\textwidth]{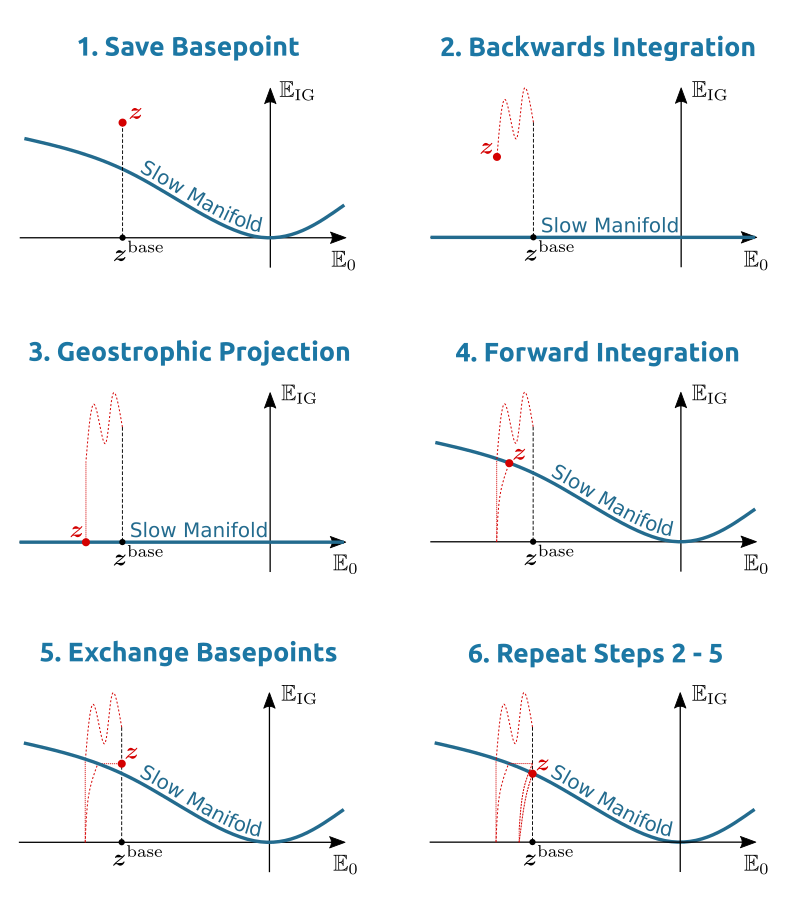}}
  \caption{
  The six steps of the OB procedure illustrated in a phase space diagram. Each panel shows the phase space at the end of corresponding OB step. The horizontal axis is the magnitude of the linear geostrophic component and the vertical axis the magnitude of the linear inertia-gravity wave component. The red dashed line is the trajectory of the state $\boldsymbol{z}$ that is projected onto the slow manifold. }
  \label{fig:OptimalBalance}
\end{figure}

When ramping an arbitrary state to the linear end, projecting onto the geostrophic eigenspace and then back to the nonlinear end, the obtained state may be balanced. However, the linear geostrophic mode of the backward-forward ramped state may not be the same as the linear geostrophic mode of the initial state, and a subsequent iteration might introduce  further drift.
As a consequence, the backward-forward ramping algorithm would be inconsistent with the NNMD method by \citet{Machenhauer:77,WarBokShe95}. 
To ensure consistency,  we follow \citet{masur2020optimal}
and replace the linear geostrophic mode of the backward-forward ramped state with the linear geostrophic mode of the initial state. Since the new state, after exchanging the linear geostrophic mode, may no longer be in balance, the whole procedure is iteratively repeated until the algorithm eventually converges. In the following, we give a step by step instruction of OB, 
illustrated in Fig.~\ref{fig:OptimalBalance}.

\subsubsection*{Step 1: Save the base point coordinate} 

 The base point coordinate $\boldsymbol{z}^{\text{base}} $ is a boundary value in time. As discussed before, the linear geostrophic mode of the initial and the balanced state should stay the same. Hence, we use the linear geostrophic mode of the initial unbalanced state $\boldsymbol z$ as the base point coordinate, which will stay constant during the iteration. Other possible choices for the base point coordinate are for example the potential vorticity, or the velocity field. However, with the geostrophic mode of $\boldsymbol z$ as the base point coordinate, the convergence of the OB procedure seems to be fastest
 \citep{masur2020optimal}.
 The first step of the OB procedure (Fig.~\ref{fig:OptimalBalance}.1) is to save the base point coordinate of the state.
\begin{equation}
    \boldsymbol{z}^{\text{base}} = {\mathsfbi P}_0 \bcdot \boldsymbol z
\end{equation}

\subsubsection*{Step 2: Backward integration to linear end}

The initial unbalanced state $\boldsymbol z$ is integrated backward in time while ramping the nonlinear term to zero
\begin{equation}
  \boldsymbol z^{(ii)} = \mathcal{R}^b (\boldsymbol z)
\end{equation}
where the superscript $(ii)$ is a label to denote the state after the second step. The backward ramping operator $\mathcal{R}^b$ is defined as
\begin{equation}
  \mathcal{R}^b (\boldsymbol z) = \int_\tau^0 \left[ \mathcal{L} \bcdot \boldsymbol z(t) + \rho(t) Ro\, \boldsymbol N(\boldsymbol z(t)) \right]dt
\end{equation}
In the numerical model, the backward integration can be achieved by switching the sign of the time step.  While  ramping down the nonlinear term, the slow manifold is deformed into the geostrophic eigenspace (Fig.~\ref{fig:OptimalBalance}.2). During the backwards integration, the state will oscillate fast 
if it is not balanced already, and follows the deformation of the slow manifold.

\subsubsection*{Step 3: Projection onto linear slow manifold}

After the backwards integration we end up at the linear end. Here, the state $\boldsymbol z^{(ii)}$ is projected onto the geostrophic eigenspace (Fig.~\ref{fig:OptimalBalance}.3)
\begin{equation}
  \boldsymbol z^{(iii)} = {\mathsfbi P}_0 \cdot \boldsymbol z^{(ii)}
\end{equation}

\subsubsection*{Step 4: Forward integration to nonlinear end}

The linear geostrophic mode $ \boldsymbol z^{(iii)} $ is integrated forward in time while ramping the nonlinear ramp factor $\rho$ back to one. The forward integration must be exactly reversed to the backward integration in step 2
\begin{equation}
    \boldsymbol z^{(iv)} = \mathcal{R}^f (\boldsymbol z^{(iii)})
\end{equation}
with the forward ramping operator $\mathcal{R}^f$
\begin{equation}
    \mathcal{R}^f (\boldsymbol z) = \int^\tau_0 \left[ \mathcal{L} \bcdot \boldsymbol z(t) + \rho(t) Ro\, \boldsymbol N(\boldsymbol z(t)) \right]dt
\end{equation}
As discussed earlier, the state will follow the deformation of the slow manifold (Fig.~\ref{fig:OptimalBalance}.4).

\subsubsection*{Step 5: Apply boundary condition at the nonlinear end}

After the fourth step, the state $\boldsymbol z^{(iv)}$ is in balance and on the slow manifold $S$ but the linear geostrophic component of $\boldsymbol z^{(iv)}$ may differ 
from the base point coordinate $\boldsymbol z^\text{base}$, which we saved in step 1.
In order to satisfy that  boundary value, we set (Fig.~\ref{fig:OptimalBalance}.5)
\begin{equation}
    \boldsymbol z^{(v)} = \boldsymbol z^{(iv)} - {\mathsfbi P}_0 \cdot \boldsymbol z^{(iv)} + \boldsymbol z^\text{base}
    \label{eq:ob5_with_geostrophic}
\end{equation}

\subsubsection*{Step 6: Repeating step 2 - step 5}

By replacing the base point coordinates in step 5, the state may no longer be in balance
and on the slow manifold $\mathcal{S}$, as illustrated 
in Fig.~\ref{fig:OptimalBalance}.5. However, the procedure has reduced the amplitude of the unbalanced component. Hence, repeating the whole procedure may further reduce the amplitude of the unbalanced component. With each iteration, the amplitude of the unbalanced component becomes smaller until it eventually converges to the optimal balanced state. A quasi convergence of this algorithm has been shown by \citet{masur2023quasi}. We denote the OB balancing operator with $m$ backward-forward iterations with
\begin{equation}
  \mathcal{B}[m, 2\rightarrow 5, {\mathsfbi{P}}_0]
  \label{eq:OB_operator}
\end{equation}
where the $2\rightarrow 5$ refers to the repetition of step 2 to 5, and the argument ${\mathsfbi{P}}_0$ denotes that we use the geostrophic projector with spectral decomposition to project onto the linear slow manifold, and to calculate the base point coordinate. In OBTA, we replace the geostrophic projector ${\mathsfbi{P}}_0$ with a new time-averaging projector.

\subsection{The optimal balance method with time-averaging (OBTA)}
\label{subsec:geo_pro_TA}

It is clear that a fundamental prerequisite for the OB method by \citet{masur2020optimal} is the ability to project onto the linear slow manifold, which in our case is the projection onto the linear geostrophic mode using the spectral projector ${\mathsfbi P}_0$, which relies in turn on a Fourier transformation. Hence, in models where Fourier transformations are inaccessable due to irregular boundaries, varying depth or other environmental conditions, or complicated due to  unstructured grids, we can not use the spectral projector. 
 To solve this problem, we replace the spectral projector ${\mathsfbi P}_0$ that is used in the OB method 
 by \citet{masur2020optimal}
 with a linear geostrophic projector ${\mathsfbi P}_T$ that does not require a Fourier transformation. 

The idea of the linear geostrophic projector ${\mathsfbi P}_T$ is based on the fact that in the linear model, the geostrophic mode is constant in time, while the inertia-gravity waves are oscillating in time. These oscillations are filtered out in an infinite cumulative time average of the linear model, 
while the geostrophic mode remains unaffected. To demonstrate this, we define the cumulative time average as
\begin{equation}
  \langle \boldsymbol{z} \rangle^T = \frac{1}{T} \int_{t_0}^{t_0+T}  \Phi_L^t \boldsymbol z \, dt
  \label{eq:time_average_single}
\end{equation}
where the evolution operator $\Phi_L^t$ only evolves the linear term of Eq.~(\ref{eq:SWeq}). Inserting the general solution of the geostrophic and of the first inertia-gravity wave mode Eq.~(\ref{eq:general_solution}) in the cumulative time average yields Eq.~(\ref{eq:single_average_error}). Note that we do not explicitly show the second inertia-gravity wave mode $\hat{\boldsymbol{z}}_2$ here, as it is similar to $\hat{\boldsymbol{z}}_1$.
\begin{equation}
  \langle \hat{\boldsymbol{z}}_0 \rangle^T = \hat{\boldsymbol{z}}_0
  \hspace{2cm}
  \langle \hat{\boldsymbol{z}}_{1} \rangle^T = \hat{\boldsymbol{z}}_{1} \frac{i}{\omega_1 T}\left( e^{-\mathrm{i}\omega_{1}T} - 1 \right) 
  \equiv \delta \hat{\boldsymbol{z}}_{1}
  \label{eq:single_average_error}
\end{equation}
where $\delta(\boldsymbol{k},T) \in \mathbb C$ is a complex damping factor. $\delta$ converges uniformly to zero for $T\rightarrow \infty$. Hence, the cumulative time average of the linear model indeed converges to the geostrophic mode. However, as the damping factor $\delta$ scales with $T^{-1}$, the time-averaging projection is computationally expensive. The computational cost can be decreased by applying the time average iteratively, e.g. by time-averaging the time averaged state
\begin{equation}
  {\mathsfbi{P}}_{T,n}^\text{con}(\hat{\boldsymbol{z}}_1) \equiv
  {\langle \dots \langle }
  \hat{\boldsymbol{z}}_1
  \underbrace{\rangle^T \dots \rangle^T}_{n-\text{times}} 
  = 
  \delta^n 
  \hat{\boldsymbol{z}}_1
\end{equation}
Hence, the error scales with $1/(\omega_1 T)^n$ for the iterative time-averaging while the error scales with $1/(n\omega_1 T)$ when doing the single time average n-times longer. Thus, when the averaging period $T$ is large enough, it is better to repeat the time-averaging rather than to increase the time-averaging period. However, when the time-averaging period is too small, for example when $T < \omega_1^{-1}$, a longer time average may be better than multiple very short ones.

The damping factor $\delta$ is not only small when the factor $1/(\omega_1 T)$ is becoming small, but also when the exponential term $e^{-i\omega_1 T}$ is approaching $1$. The exponential term is equal to $1$ when the averaging period is equal to the period of the inertia-gravity wave, or a multiple of it. This implies, that all waves, whose period match with the averaging period are filtered out exactly. This property can be used to further improve the convergence rate of the time-averaging procedure: instead of doing all time averages with the same time-averaging period, we alter the averaging period for each individual time average. We propose to use equidistantly spaced averaging periods between an inertial period $T=T_f$ and half an inertial period
\begin{equation}
  {\mathsfbi P}_{T,n}^\text{equi} (\boldsymbol z) = \langle \dots \langle \boldsymbol z \rangle^{T_1} \dots \rangle^{T_n}
  \hspace{1cm} \text{with} \hspace{1cm}
  T_i = \frac{2n + 1 - i}{2n} T
  \label{eq:geoproj_time_average}
\end{equation}
The averaging with $T_i=T/2$ is excluded since waves with that period are already filtered out by the averaging with $T_i=T$. In the experiment discussed below, we compare the method with equidistantly spaced averaging periods (${\mathsfbi P}_{T,n}^\text{equi}$), hereafter referred to as the equidistant time chunk method, with the method using constant averaging periods (${\mathsfbi P}_{T,n}^\text{con}$), hereafter referred to as the constant time chunk method. The general notation ${\mathsfbi{P}}_T$ stands for ${\mathsfbi P}_{T,n}^\text{con}$ or ${\mathsfbi P}_{T,n}^\text{equi}$.

Given that the geostrophic projection using time-averaging is not exact, the question raises on how this averaging error affects the accuracy of balancing procedure. To estimate the accuracy of the OBTA procedure, we decompose the initial condition in the linear geostrophic mode ${\boldsymbol{z}}_0$, the slaved mode ${\boldsymbol{z}}_s$, and the residual ${\boldsymbol{z}}_r$
\begin{equation}
  {\boldsymbol{z}} = {\boldsymbol{z}}_0 + {\boldsymbol{z}}_s + {\boldsymbol{z}}_r
\end{equation}
where ${\boldsymbol{z}}_0 + {\boldsymbol{z}}_s$ is the balanced state that we want to obtain with the balancing procedure. To compare OB with OBTA we denote states in the OBTA procedure with a tilde, e.g. the base point coordinate of OBTA is given by
\begin{equation}
  {\boldsymbol{\Tilde z}}^\text{base} = {\boldsymbol{z}}_0 + \delta {\boldsymbol{z}}_s + \delta {\boldsymbol{z}}_r
  \label{eq:base_point_error}
\end{equation}

where $\delta$ is the time-averaging error Eq.~(\ref{eq:single_average_error}). The error of the base point coordinate is hence given by the time-averaging error of the slave mode and the residual. Since OB restores the base point coordinate after each iteration step, we may conclude that the accuracy of OBTA is bounded by the accuracy of the time-averaging. However, with another small modification of the method we can eliminate this limitation, such that OBTA converges to OB independently of the time-averaging error $\delta$. Instead of repeating step 2-5 in the iteration phase, we repeat step 1-5 in OBTA. This modification means that we recalculate the base point coordinate before each iteration step. To demonstrate why this recalculation is useful, we calculate the OBTA error after  step 5
\begin{equation}
  \boldsymbol{\Tilde z}^{(v)} = 
  \boldsymbol{z}^{(v)} + \delta \
  \textit{O} (\boldsymbol{z}_r)
  \label{eq:OBTAerror}
\end{equation}
where $\boldsymbol{z}^{v}$ is the OB state after step 5. 
More details on this calculation and the underlying assumption are given in the Appendix \ref{appendix:OBTA_error}. Surprisingly, the additional balancing error that is caused by the time-averaging only depends on the magnitude of the residual state, but not on the magnitude of the slave mode. The residual of the obtained state $\boldsymbol{\Tilde z}^{(v)}$ is by the factor $\delta$ smaller as the residual of the initial condition. Hence, with each further repetition of step 1-5, the residual gets damped by another factor $\delta$, until it eventually converges to zero. The boundary condition that the geostrophic modes of the initial condition and the balanced state are the same remains intact since the time-averaging error projects onto the gravity-inertia wave mode (see Appendix \ref{appendix:OBTA_error}).

 In summary, OBTA introduces two differences from OB. First, the spectral geostrophic projector ${\mathsfbi P}_0$ is replaced with the time-averaging projector ${\mathsfbi P}_T$. And second, while in the OB procedure the base point coordinate is calculated only once in the beginning, in the OBTA procedure, the base point coordinate is recalculated before each iteration step. Using the notation from Eq. (\ref{eq:OB_operator}), the OBTA balancing operator is given by:
\begin{equation}
  \mathcal{B}[m, 1\rightarrow 5, {\mathsfbi{P}}_T]
  \label{eq:OBTA_operator}
\end{equation}

\section{Numerical Experiments}
\label{sec:numerical_experiments}
We perform three different numerical experiments: In the first experiment, we investigate whether the geostrophic projection using time-averaging (${\mathsfbi{P}}_T$) converges to the geostrophic projection using spectral decomposition (${\mathsfbi{P}}_0$). Further, we compare the convergence rate of the different time-averaging setups, i.e. the constant time chunk setup with the equidistant time chunk setup. In the second experiment, we investigate whether OBTA converges to OB when either the number of iteration steps is increased or the projection error of the time-averaging procedure is decreased. In the third experiment, we apply OBTA in a classical application case for balancing methods: the estimation of spontaneous wave emission using the diagnosed imbalance. The numerical models and initial conditions are introduced in Section \ref{sec:numerical_models}.

\subsection{Geostrophic projection error with time-averaging}
We define the geostrophic projection error $\delta_\text{proj}$ as the norm of difference between the geostrophic projection using time-averaging and the geostrophic projection using spectral decomposition:
\begin{equation}
  \delta_\text{proj}(\boldsymbol{z}) = 
  d\left({\mathsfbi{P}_0} (\boldsymbol z), {\mathsfbi{P}}_T (\boldsymbol z) \right)
  \hspace{1cm} \text{with} \hspace{1cm}
  d(\boldsymbol{z}_1, \boldsymbol{z}_2) = 2 \frac{|| \boldsymbol{z}_1 - \boldsymbol{z}_2||}{|| \boldsymbol{z}_1 || + || \boldsymbol{z}_2 ||}
  \label{eq:diagnosed_imbalance}
\end{equation}
where $d(\boldsymbol{z}_1, \boldsymbol{z}_2)$ denotes the norm of difference between two states, and $|| \cdot ||$ marks the $L^2$ norm.
The vector $\boldsymbol z$ is given by Eq.~(\ref{eq2})
for the shallow water model and accordingly for the
non-hydrostatic model (see, e.g. \citet{Chouksey:22}).
The goal of this experiment is to find out which time-averaging setup converges fastest to the geostrophic mode.
We measure the convergence speed with the computational cost which is proportional to the total averaging time $T_\text{tot}$. The total averaging time for the constant time chunk setup is given by:
\begin{equation}
  T_\text{tot}\left({\mathsfbi{P}}_{T,n}^\text{con}\right) = \sum_{i=1}^n T = nT
\end{equation}
and for the equidistant time chunk setup by:
\begin{equation}
  T_\text{tot}\left({\mathsfbi{P}}_{T,n}^\text{equi}\right) = \sum_{i=1}^n \frac{2n+1-i}{2n}T = \frac{3n+1}{4}T
\end{equation}
The best time-averaging setup is the setup which has the smallest projection error for a given total averaging time.

\subsection{Deviation between OBTA and OB balanced states}
In this experiment we want to test whether OBTA converges to OB when either increasing the number of iteration steps, or the accuracy of the time-averaging setup. To achieve this, we define the deviation of a balanced state $\mathcal{B}[\dots](\boldsymbol{z})$ to the OB balanced state $\boldsymbol{z}_\text{ref}$ as:
\begin{equation}
  \epsilon_\text{dev}(\boldsymbol{z}) = d(\boldsymbol{z}_\text{ref}, \mathcal{B}[\dots](\boldsymbol{z}))
  \hspace{2cm} \text{with} \hspace{0.5cm}
  \boldsymbol{z}_\text{ref} = \mathcal{B}[10,2\rightarrow 5, {\mathsfbi{P}}_0](\boldsymbol{z})
  \label{eq:balancing_error}
\end{equation}
where $d(\dots)$ denotes the norm of difference. Note that we use OB with $10$ backward-forward iterations as our reference balancing method because in most of our experiments, OB seems to be converged after less than 10 iterations.  

\subsection{Diagnosed imbalance}
With the last two experiments described in the previous subsections, we aim to estimate and compare the accuracy of the balancing methods. 
The deviation term $\epsilon_\text{dev}$ is an inadequate measure for the balancing accuracy.
This inadequacy arises from two main reasons: First, the deviation term assumes that the OB balanced state $\boldsymbol{z}_\text{ref}$ is the best balanced state. 
Hence a more accurate balanced state than $\boldsymbol{z}_\text{ref}$ would falsely be considered as a less accurate balanced state.
Second, it is unclear whether the balancing problem is unique, i.e. whether there exists only one balanced state for a given initial condition.
A more appropriate measure for the balancing accuracy is the diagnosed imbalance.
The diagnosed imbalance tests whether a trajectory of an initially balanced state remains balanced, i.e. to what extent
\begin{equation}
    \mathcal{B} \Phi^t \mathcal{B} \boldsymbol z = \Phi^t \mathcal{B} \boldsymbol z
    \label{eq:invariantbalancing}
\end{equation}
holds for a given state $\boldsymbol{z}$ and balancing method $\mathcal{B}$. We calculate the distance between the trajectory, and its projection onto the balanced mode with the norm of difference:
\begin{equation}
    Im(\boldsymbol{z}, t) = 
    d(\Phi^t \mathcal{B} \boldsymbol z, \mathcal{B} \Phi^t \mathcal{B} \boldsymbol z)
\end{equation}
where $Im(\boldsymbol{z}, t)$ is called the diagnosed imbalance. The diagnosed imbalance can be directly calculated for any given state $\boldsymbol z$ and a given diagnosing period $t$. 

A large value of the diagnosed imbalance could potentially mean two things: Either there is spontaneous wave emission, where the balanced component emits the fast waves. 
Or the balancing operator is inaccurate, i.e. the initial state is not exactly balanced. 
Hence, the diagnosed imbalance cannot not be interpreted as an absolute measurement of the inaccuracy of a balancing operator. 
However, it can be used to compare the inaccuracies between different balancing methods. 
We consider a balancing operator that obtains the smallest diagnosed imbalances as the most accurate balancing operator. The concept of diagnosed imbalance has proven useful before and is for example used by \citet{masur2020optimal,chouksey2023comparison}.

\section{Numerical models}
\label{sec:numerical_models}

In this work we use the FRIDOM 
(Framework for Idealized Ocean Models) model,
which re-implements the shallow water model 
and the three-dimensional non-hydrostatic model used
in \citet{Eden:2019gravity}, \citet{chouksey2023comparison}, etc 
in python and which is optimized for GPUs. 
The source code of the model can be found at \url{https://github.com/Gordi42/FRIDOM}. 
The discretisation of the shallow water model is identical to the finite difference model of \citet{chouksey2023comparison}, with a 
 double periodic domain of size $2\upi \times 2\upi$,
 which is resolved with  $511 \times 511$ grid points. 
The discretisation of the non-hydrostatic model is identical to 
the one used in \citet{Eden:2019gravity} and \citet{Chouksey:22}. 
We use a triple periodic domain of size, $4\times 4\times 1$  
with a resolution
of $255 \times 255 \times 31$ grid points
in $x$, $y$, and $z$ direction respectively.
Increasing the resolution to
$511 \times 511 \times 63$ grid points for some for the experiments described below did not change 
significantly the results.
There is explicit damping 
in the non-hydrostatic model by
biharmonic mixing and friction with coefficients
of $0.1 Ro \Delta x^3$ and  $0.1 Ro \Delta z^3$ 
in lateral and vertical direction, respectively,
but no such damping in the shallow water model.
For both models, 
we use a third-order Adam Bashforth time-stepping scheme. 
The backward integration necessary for OB and OBTA
is realised by switching the sign of the time step $\Delta t$. 
At the beginning of the time integration, when there is not yet enough information about the system's past, a time-stepping method with the next lower order is used. For example, time step $n=1$ is calculated with an Euler forward stepping, $n=2$ with the second-order Adam-Bashforth time stepping, and $n \geq 3$ with the third-order Adam-Bashforth time stepping. At the start of the backward and forward ramping during OB and OBTA, the same time stepping initialization is used. 

In the shallow water model we test two different configurations, namely an unstable free jet, and a random phase state with a prescribed energy spectra. In the non-hydrostatic model we test an unstable free jet with horizontal and vertical shear velocities.

\subsection{Unstable free jets in the shallow water model}
\label{subsec:SW_ini_jet}

\begin{figure}
  \centerline{\includegraphics[width=0.8\linewidth]{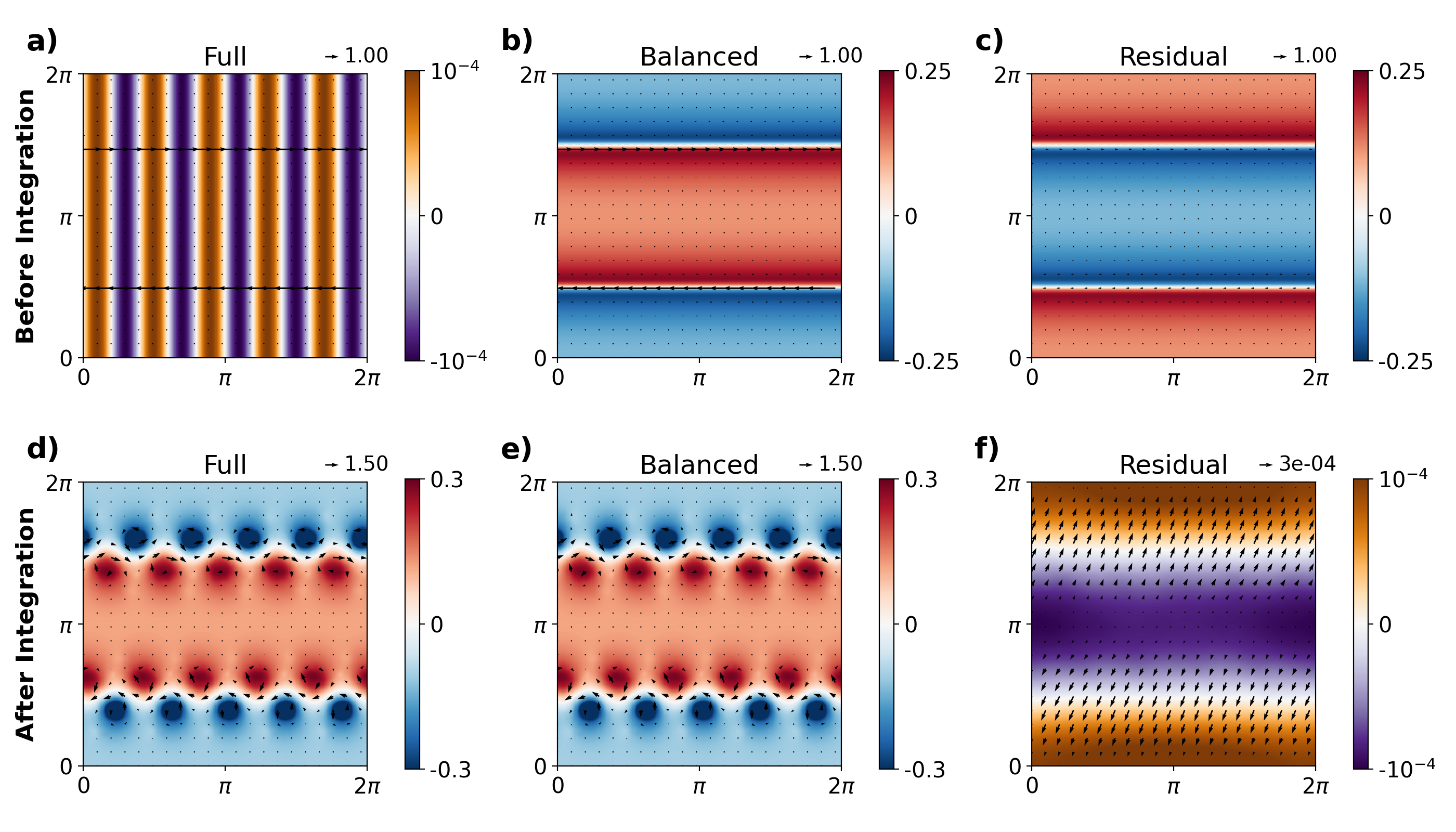}}
  \caption{Evolution of the unstable free jets at $Ro=0.3$ in the shallow water model.
  The top row shows the free-jet state configuration (a) and its decomposition in balanced (b) and residual (c) using OB. The balanced state (b) is set as the initial condition in the model.
  The bottom row (d) shows the obtained state after a model integration of $t=10/Ro$ and its decomposition in balanced (e) and residual(f). The balanced state are obtained from $\mathcal{B}[10,2\rightarrow 5, {\bf P}_0]$. The arrows show the velocity field, and the colors represent the layer thickness $h$. For fields with very small layer thicknesses (i.e. a,f), a different colormap is used than for those with large layer thicknesses.}
  \label{fig:jet_fields}
\end{figure}

The first configuration of the shallow water model is given by two opposing jets
as in \citet{chouksey2023comparison}; 
an eastward-flowing jet in the northern half of the domain and a westward-flowing jet in the southern half. A small sinusoidal perturbation with a wavelength that fits five times in the domain is added to the height field (Fig.~\ref{fig:jet_fields} a), to start
a lateral shear instability.
For more details on the configuration we refer the reader to \citet{chouksey2023comparison}. In contrast to \citet{chouksey2023comparison}, we do not project the obtained state onto its geostrophic mode, which means that the testing state is not purely geostrophic. We made this modification to challenge the time-averaging procedure in OBTA, i.e. it has to filter out the non-geostrophic component.

For the calculation of the diagnosed imbalance, the testing state (Fig.~\ref{fig:jet_fields} a) is balanced using the balancing method that we want to test. The obtained balanced state (Fig.~\ref{fig:jet_fields} b) is initialized in the shallow water model and integrated forward in time. During the forward integration, the jets start to meander and form  cyclonic and anticyclonic eddies. This evolved state is shown in Fig.~\ref{fig:jet_fields} d) for $Ro=0.3$ and $t=Ro/10$.
Finally, the evolved state is balanced again, the norm of difference between the evolved state and the balanced evolved state is diagnosed imbalance.
We tested the diagnosed imbalance as a function of diagnosing time $t$ and found that it grows exponentially during the growth of eddies, and remains approximately constant 
for times larger than $Ro/10$ (not shown).
As discussed in Section \ref{sec:OB}, the accuracy of OB depends on the choice of the ramp period $\tau$, sensitivity experiments with different ramp periods and  different Rossby numbers yield the smallest imbalances for $\tau=3/Ro$. 

\subsubsection{Random phases in the shallow water model}

\begin{figure}
  \centerline{\includegraphics[width=0.8\linewidth]{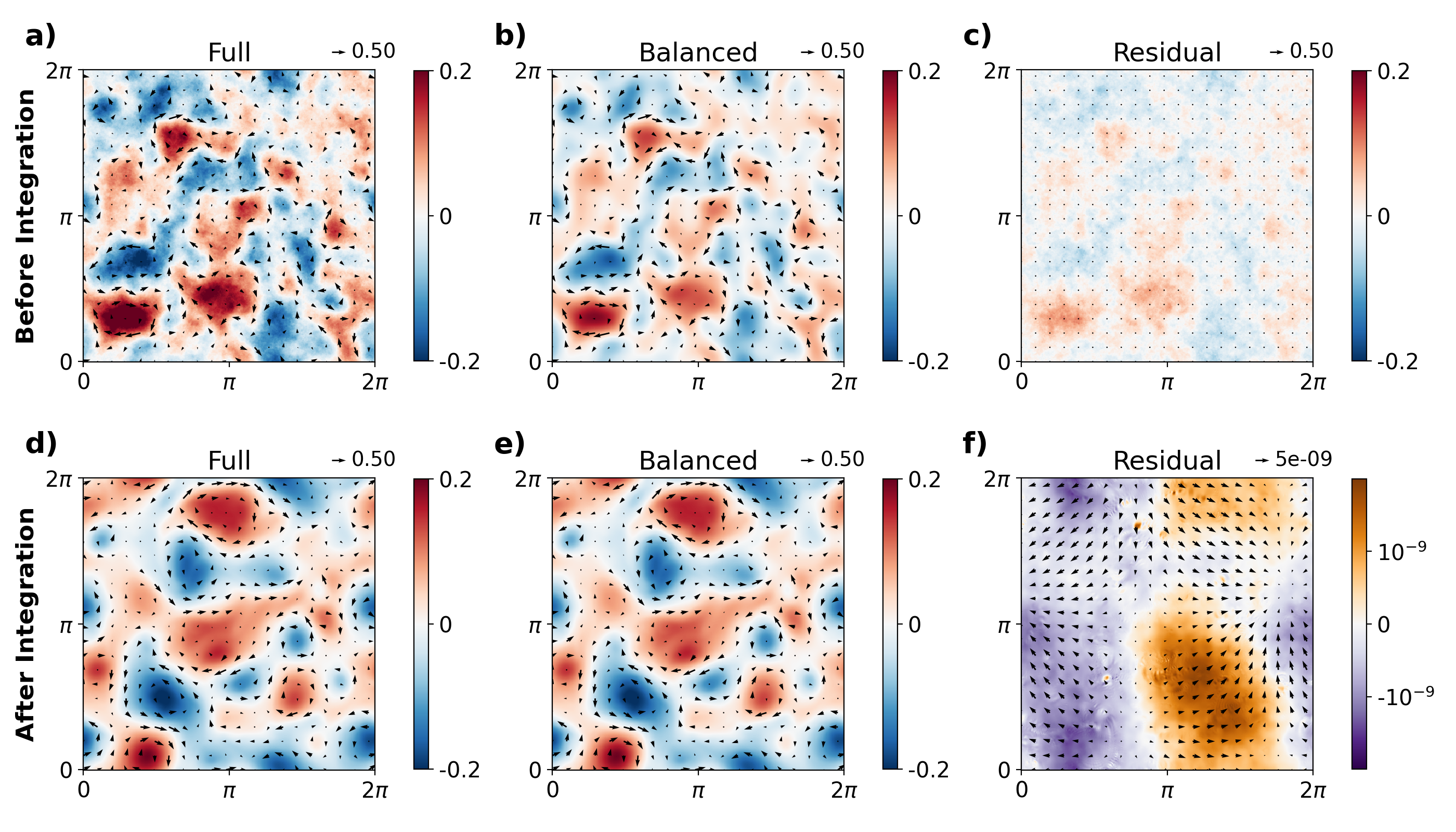}}
  \caption{Same as Fig.~\ref{fig:jet_fields}, but for the random phase case.}
  \label{fig:ran_fields}
\end{figure}

The second configuration that we use for the shallow water model are
generated from prescribed kinetic energy spectra with random phases. We prescribe different spectras for the geostrophic component and for the wave component.
The spectrum of the geostrophic mode has initially 
a maximum at $k=6$ and slopes for small wavelength with $k^{-6}$, again similar to \citet{chouksey2023comparison}.
We multiply each Fourier coefficient from the spectrum
with a random complex number taken from a uniform distribution to randomize the phase.
In addition to the geostrophic spectrum, 
we add a wave mode with a spectral kinetic energy spectrum that scales with $\omega^{-2}$ 
motivated by the spectral model introduced by \citet{Garrett:75}. 
We add the wave component  $\hat{\boldsymbol{z}}_\text{IG}$ with
\begin{equation}
    \hat{\boldsymbol{z}}_\text{IG}(\boldsymbol{k}) = \omega^{-1} 
    (r_1 (\boldsymbol{k}) \boldsymbol{q}_1 + r_2(\boldsymbol{k}) \boldsymbol{q}_2)
\end{equation}
to the geostrophic spectrum, 
where $\boldsymbol{q}_i$ are the eigenvectors and $r_i(\boldsymbol{k})$ are random complex numbers taken from an uniform distribution.
The magnitude of the wave component is scaled such that the maximum layer thickness is 0.1, while the maximum layer thickness of the geostrophic mode is scaled to 0.2. Hence, the magnitude of the wave component is half as large as that of the geostrophic mode. 

Fig.~\ref{fig:ran_fields} shows the resulting
random phase testing state and the decomposition into balanced and wave components using the OB.
For the random phase case we use a ramp period of $10/Ro$, which yields smalles diagnosed imbalances. 
Evolving the balanced field for $Ro=0.3$ and a diagnosing period of $10/Ro$ yields the evolved state shown in Fig.~\ref{fig:ran_fields} d). The power spectra (not shown here) of the evolved state has shifted towards larger wave numbers, consistent with the inverse energy cascade of geostrophic flow.

\subsubsection{Unstable free jets in the non-hydrostatic model}

\begin{figure}[t]
  \centerline{ \includegraphics[width=0.8\linewidth]{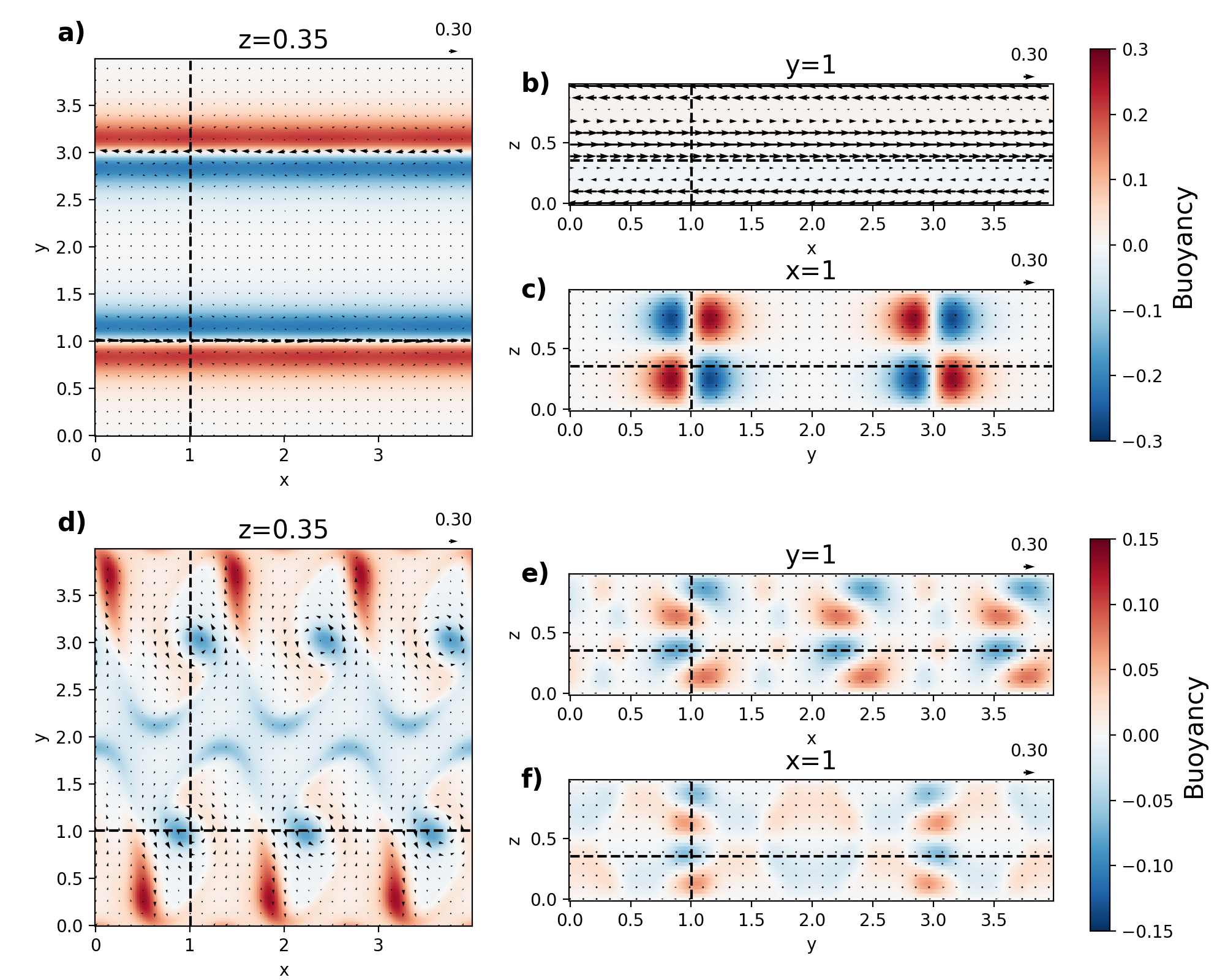}}
  \caption{Buoyancy $b$ of the initial balanced jet (a-c), and after an integration period of $10/Ro$ for a Rossby number of $0.1$ (d-f)
   in the non-hydrostatic model. Panels a) and d) show a top section taken at $z=0$, b) and e) show a side section in the x,z plane at $y=1$, and c) and f) show a side section in the y,z plane at $x=1$. The dashed lines indicate the position of the sections.}
  \label{fig:jet3d_fields}
\end{figure}

For the non-hydrostatic model we choose only
one configuration, similar to the one in \citet{Chouksey:22}. It is also given by unstable 
east- and westward jets, for which 
the initial u-velocity is given by
\begin{equation}
  u = \left(\exp{-\frac{(y-3)^2}{0.16^2}} - \exp{-\frac{(y-1)^2}{0.16^2}} \right) \cos{2 \pi z}
\end{equation}
The vertical structure of the jets corresponds to
the first baroclinic linear eigenmode.
To seed the lateral shear instability of the jets,  
a perturbation given by the geostrophic eigenvector for $\boldsymbol{k} = (6 \pi / 4, 0, 0)$ is added to the jet
structure and scaled such that the maximum velocity of the perturbation is $0.05$. Fig.~\ref{fig:jet3d_fields} show the initial balanced state and the evolved state in terms of the buoyancy for $Ro=0.1$.



\section{Results}\label{sec:results}

In OBTA, the geostrophic projection is obtained with a time-averaging procedure, as introduced in Section \ref{subsec:geo_pro_TA}. 
To understand the accuracy of OBTA, we first directly assess the error using the time-averaging operator 
 ${\mathsfbi P}_T$
instead of the spectral operator ${\mathsfbi P}_0$ for the geostrophic projections. We perform our experiments in the different configurations of the shallow water model in Section \ref{sec:geo_proj_error}. 
Subsequently, in Section \ref{sec:res_obta_vs_ob} we show that the deviation between the OBTA balanced state and the OB balanced state becomes exponentially small.
Finally, in Section \ref{sec:res_diagnosed_imbalance}
we compare the diagnose imbalance obtained with OBTA and OB in the 2D shallow water model and in the 3D non-hydrostatic model.

\subsection{Geostrophic projection error with time-averaging}
\label{sec:geo_proj_error}

\begin{figure}
  \centerline{\includegraphics[width=0.8\linewidth]{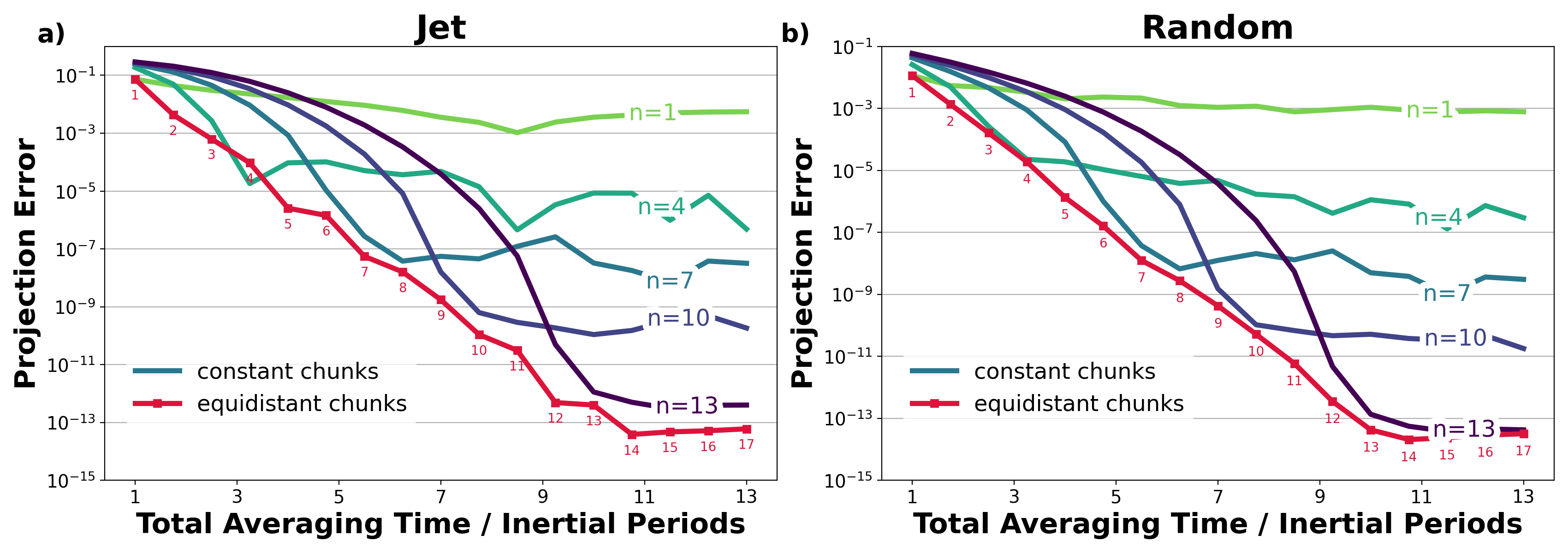}}
  \caption{The geostrophic projection error $\delta_\text{proj}$ for different time-averaging methods in the shallow water model.
  The red line shows the projection error for the averaging setup with equidistant time chunks (Eq. \ref{eq:geoproj_time_average}), where the label below each square denotes the number of time chunks $n$. The remaining lines show the projection error obtained from the constant time chunk setup, where each time average has the same averaging period and is repeated $n$ times.}
  \label{fig:AccuracyGeostrophicProjection}
\end{figure}

Fig.~\ref{fig:AccuracyGeostrophicProjection} shows
the accuracy of ${\mathsfbi P}_T$ in the two different shallow water model
configurations and the different methods as a function of the total averaging time. Since the total averaging time is a function of both, the chunk size of the individual time averages $T$, and the number of time chunks $n$, we keep one parameter constant while varying the other one. For the averaging setup with constant time chunks, we keep $n$ constant, while varying the chunk size. For the equidistant time chunk setup, the time chunk sizes are always equidistantly spaced between a full and a half inertial period, while the number of time chunks is increased from $1$ to $17$.

We start by considering the case of a single time average with no chunks, $n=1$. Theoretically, the error should be reduced by half when doubling the averaging time (Eq.~\ref{eq:single_average_error}). This behaviour can be seen in Fig.~\ref{fig:AccuracyGeostrophicProjection} b), where the curve of the projection error becomes flatter for longer averaging times. 
However, in Fig.~\ref{fig:AccuracyGeostrophicProjection} a), the projection error gets smaller until an total averaging time of roughly 9 inertial periods and then increases again. 
But  longer time-averaging periods than shown in Fig.~\ref{fig:AccuracyGeostrophicProjection} (not shown here) reveal that the observed minimum is a local minimum, similar as for $n=4$, and the projection error continues to decrease when increasing the time-averaging period.


When increasing the number of time chunks, two different issues emerge in the behavior of projection errors for varying total time-averaging periods. For short averaging periods, the projection error becomes larger when increasing $n$. In contrast, for long averaging periods, the error becomes smaller. For $n=13$, the error is more than 10 orders of magnitude smaller than for $n=1$. We search for the averaging setup that has the smallest projection error for a given total averaging time. In other words, we compare all methods with the same computational cost, and pick the one with the smallest projection error. This best averaging setup is for both initial conditions and over all total averaging periods, the equidistant time chunk setup. As discussed in Sec.~\ref{subsec:geo_pro_TA}, this is due to the fact that frequencies with periods close to the averaging period are most efficiently filtered. In the equidistant time chunk setup, many different averaging periods are used, and hence a broader spectrum of frequency is filtered out more efficiently.

For all averaging setups, the projection error does not decrease 
beyond a certain threshold value. 
This plateau may be attributed to numerical errors, as theoretically, 
the projection error should converge to zero. 
Since this limit is a lot smaller than the imbalances obtained with OB (see below), 
this numerical limit seems to be of no importance for practical applications. 
For all subsequent experiments, we adopt the equidistant time chunk method.

\subsection{Deviation between OBTA and OB balanced states}
\label{sec:res_obta_vs_ob}

\begin{figure}
  \centerline{\includegraphics[width=0.8\linewidth]{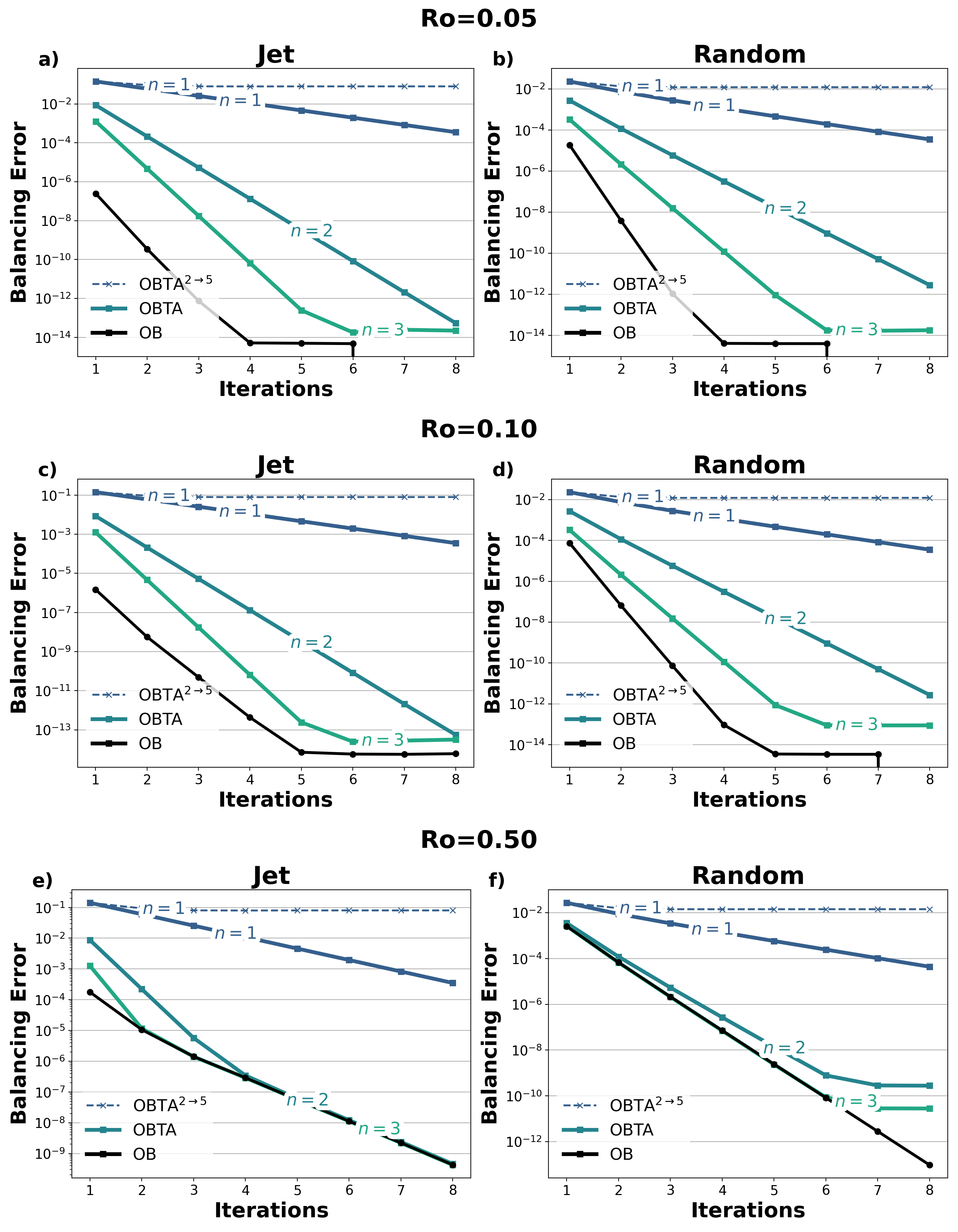}}
  \caption{The balancing error $\epsilon_\text{dev}$ (Eq.~\ref{eq:balancing_error}) as a function of the backward forward iteration $m$. The black line shows the balancing error of OB with spectral projection, the solid colored line show the OBTA error with 1, 2, and 3 equidistant time chunks (${\mathsfbi P}^\text{equi}_{n,T_f}$), and the dashed line show the OBTA error with 1 equidistant time chunk, and the repetition of step 2 to 5, instead of 1 to 5, i.e. without the recalculation of the base point coordinate. Shown are results for the Rossby numbers $0.05$ (top), $0.1$ (middle), and $0.5$ (bottom), for the jet (left) and random phase (right) initial conditions in the shallow water model.}
  \label{fig:obta_convergence}
\end{figure}

In contrast to the linear projection onto the geostrophic mode, for the nonlinear balancing problem no exact solution is known. As a consequence, the balancing error cannot be defined as the deviation from an exact solution. Instead, in Eq.~\ref{eq:balancing_error}, we define the balancing error $\epsilon_\text{dev}$ as the deviation to the OB balanced state after the 10th iteration.
In Fig.~\ref{fig:obta_convergence}, the balancing error is shown for the two shallow water configurations as a function of the number of backward-forward iterations for the different balancing methods. For OB ($\mathcal{B}[m, 2\rightarrow 5, {\mathsfbi P}_0]$), the balancing error is zero at the 10th iteration by definition. However, for $Ro=0.1$ and $Ro=0.05$, the OB balancing error stops decreasing after less than 10 iterations and remains on a constant plateau. The magnitude of this plateau is similar as for the geostrophic projection error in Fig.~\ref{fig:AccuracyGeostrophicProjection}. We conclude that the limits of numerical precision are indeed reached at this plateau.

Before the balancing error reaches this plateau, the error decreases exponentially with the number of iterations. We refer to the slope of the exponential decrease as the rate of convergence. The rate of convergence is larger, i.e. faster convergence, for smaller Rossby numbers. This relationship between the Rossby number and the convergence rate is evident for all model configurations. We speculate that this Rossby number - convergence rate relationship is linked to the amplitude of (very weak) spontaneous wave emission. 
When spontaneous wave emission is present, the assumption that the nonlinear balanced state is adiabatically transformed to the linear geostrophic mode during backward ramping is no longer satisfied. Violating the the assumption of adiabatic transformation may effects the convergence rate, with slower convergence for stronger wave emissions. The wave emission is stronger for larger Rossby numbers (see below) and hence, the convergence rate is smaller for larger Rossby numbers.

Although both model configuration reach the plateau of numerical precision at the same iteration step ($m=4$ for $Ro=0.05$ and $m=5$ for $Ro=0.1$), the convergence rate is much larger for the random phase case than for the unstable jet case. The plateau of numerical precision is reached at the same iteration step because the first balancing error (at iteration $m=1$) is multiple orders of magnitude smaller for the unstable jet case. 
The differences in the first balancing error may be related to the magnitude of the slave mode, which is two to three orders of magnitudes smaller for the unstable jet than for the random phase case. The larger convergence rate of the random phase case is likely related to the local Rossby number, which can be different from the global parameter $Ro$. The local Rossby number is given by $Ro\,\zeta/f$, where $\zeta$ denotes the horizontal vorticity of the flow field. For the random phase case, the maximum local Rossby number is about three times smaller than the maximum local Rossby number in the unstable jet case. Hence, the observation that the random phase case converges faster than the jet case agrees with the observation that the convergence is faster for smaller Rossby numbers.

We now consider the balancing error of OBTA, where the spectral geostrophic projection ${\mathsfbi P}_0$ is replaced with the time-averaging operator ${\mathsfbi P}^\text{equi}_{T_f,n}$. The balancing error for $n=1,2,3$ is shown in Fig.~\ref{fig:obta_convergence}. For all experiments, the same pattern emerges: Similar as the OB balancing error, the OBTA balancing error decays exponentially with the number of iteration steps. Thereby, the rate of convergence depends on the number of time-averaging chunks $n$, with faster convergence for larger $n$. For larger $n$, the geostrophic projection with time-averaging becomes more accurate (Fig.~\ref{fig:AccuracyGeostrophicProjection}). In Section \ref{sec:geo_proj_error}, we analytically linked the accuracy of the geostrophic projection to the convergence rate of OBTA: After each iteration, the obta error gets multiplied with the time-averaging damping factor $\delta$ (Eq.~\ref{eq:OBTAerror}), which is smaller for more accurate time-averaging setups. Hence, the findings in our experiments match well with our analytical considerations.

In Fig.~\ref{fig:obta_convergence} we also show the OBTA balancing error when step 2 to 5 are repeated instead of step 1 to 5, i.e. the base point coordinate gets not recalculated after each iteration. For this setup, the balancing error no longer decreases when increasing the number of iteration steps. This is due to the calculation of the base point coordinate. When calculating the base point coordinate with a time-averaging operator, the base point coordinate has an error that consists of the averaging error of the slave mode $\boldsymbol{z}_s$ and of the residual $\boldsymbol{z}_r$ (Eq.~\ref{eq:base_point_error}). The averaging error of the slave mode cuts out in the fifth OB step since the slave mode at step 1 and at step 4 is approximately the same (see \ref{appendix:OBTA_error}). However, the averaging error of the residual $\boldsymbol{z}_r$ remains important. The residual becomes smaller after each iteration step, and hence, the averaging error in the calculation of the base point coordinate would also get smaller after each iteration step. However, if the base point coordinate is not updated, the averaging error of the initial residual gets restored after each iteration step. Hence, the balancing error does not decrease with the iteration step when repeating step 2-5 instead of step 1-5. 

The analytical consideration suggest that the OBTA balanced state converge precisely to the OB balanced state when increasing the number of iterations. However, this is not the case in our experiments. After a certain iteration number, the difference between both states does not get smaller and remains on a constant plateau. This is most clearly visible for the random phase case and a Rossby number of 0.5. This discrepancy between theory and experiments may either be attributed to numerical errors, or that the underlying assumptions in the theory are violated. To find out the reason for this discrepancy, further experiments are required. However, as the difference between OB and OBTA is firstly very small, and secondly the difference decreases when the accuracy of the time-averaging operator is increased, we will not further investigate the cause of the discrepancy here. 
For practical applications, as for example diagnosing imbalances, such small differences in the balanced state have no effect on the result. 

\subsection{Diagnosed imbalance}
\label{sec:res_diagnosed_imbalance}

\begin{figure}
  \centering
  \centerline{\includegraphics[width=0.8\linewidth]{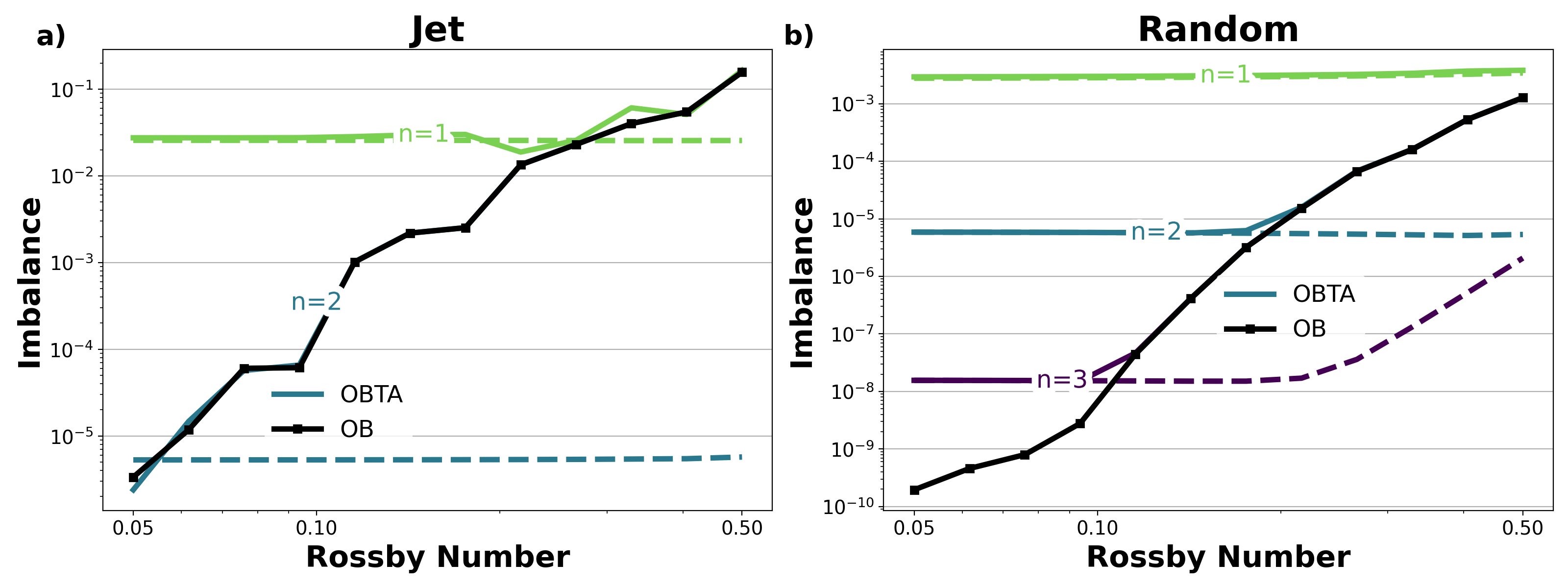}}
  \caption{The diagnosed imbalance $Im$ as a function of the Rossby number $Ro$ for the shallow water configurations. The black solid line show the OB imbalance, and the colored solid lines the OBTA imbalance with 1, 2, and 3 equidistant time chunks. The dashed coloured line show the balancing error $\epsilon_\text{dev}$ (Eq.~\ref{eq:balancing_error}) of the initial condition.}
  \label{fig:diag_imbalance_sw}
\end{figure}

The difference between OB and OBTA balanced state gets very small when either increasing the number of backward-forward iterations $m$, or the number of time-averaging chunks $n$. 
As a consequence, the differences in diagnosed imbalance between both methods should also vanish. 
However, what changes when stopping the backward-forward iterations before convergence, i.e. with a non-zero balancing error?
In this section, we always stop the backward-forward iterations after $m=3$ iterations, and compare the diagnosed imbalance between OB and OBTA.
In Fig.~\ref{fig:diag_imbalance_sw}, the diagnosed imbalance is shown as a function of the Rossby number for the shallow water model configurations. 
The diagnosed imbalance obtained with OB (black solid line) scales exponential with the Rossby number, consistent with the findings of \citet{chouksey2023comparison}.
At similar Rossby numbers, the diagnosed imbalance in the random phase case is multiple orders of magnitude smaller than in the unstable jet case. 
This difference between both configurations is likely related to the maximum local Rossby number, which is three to four times
smaller in the random phase case than in the unstable jet case.

We now consider the diagnosed imbalance obtained with OBTA for $n=2$ equidistant time chunks in the random phase case (Fig.~\ref{fig:diag_imbalance_sw}). 
For large Rossby numbers ($Ro>0.2$), diagnosed imbalances obtained with OB and OBTA are the same, however, for smaller Rossby numbers, the OB and OBTA imbalances differ from each other: 
While the OB imbalance gets smaller for decreasing Rossby number, the OBTA imbalance reaches a constant plateau and does not get smaller than a certain value. 
A similar behaviour is observed in the unstable jet case and for different $n$, where the magnitude of the plateau is smaller for larger $n$.
To understand the reason for this plateau, we also plotted the OBTA balancing error $\epsilon_\text{dev}$ of the initial condition 
(Fig.~\ref{fig:diag_imbalance_sw}, dashed lines).
The OBTA balancing error of the initial condition perfectly matches the plateau in the diagnosed imbalance. 
If the OB imbalance is smaller then the OBTA balancing error, the OBTA imbalance is given by the OBTA balancing error. 
In the opposite case, when the OB imbalance is larger than the OBTA balancing error, the OBTA imbalance is given by the OB imbalance.
Hence, to detect imbalances of a certain magnitude, the balancing inaccuracy must be slightly smaller than the magnitude of these imbalances.

One might wonder why, in most scenarios, OBTA balancing error is independent of the Rossby number, while in the specific case of the random phase initial condition with $n=3$ equistant time chunks, the OBTA balancing error depends on the Rossby number.
In the case when the OBTA balancing error depends on the Rossby number, the OBTA balancing error is given by the OB balancing error (see Fig.~\ref{fig:obta_convergence}f for $m=3$ iterations), which depends on the Rossby number. 
In all other cases, the OBTA balancing error is given by the time-averaging error, which is independent of the Rossby number.

\subsection{Diagnosed imbalance in the non-hydrostatic model}

\begin{figure}
  \centering
  \centerline{\includegraphics[width=0.5\linewidth]{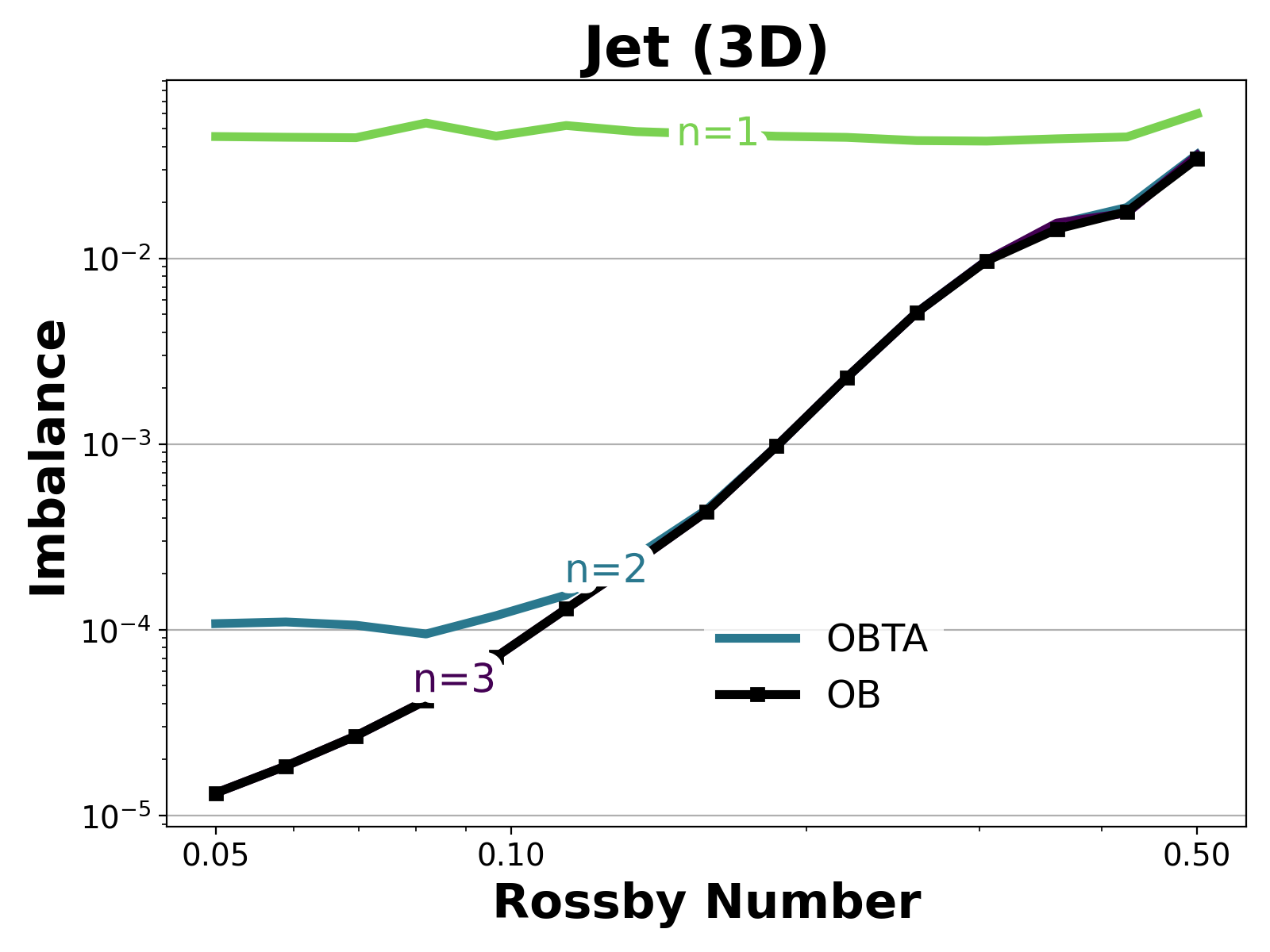}}
  \caption{Same as in Fig.~\ref{fig:diag_imbalance_sw}, but for the unstable jet in the non-hydrostatic model.}
  \label{fig:diag_imbalance_nh}
\end{figure}

In Fig.~\ref{fig:diag_imbalance_nh}, the diagnosed imbalance is shown for the unstable jet in the non-hydrostatic model. 
In the non-hydrostatic model, the diagnosed imbalance obtained with OB is of comparable magnitude as the diagnosed imbalance in the unstable jet configuration in the shallow water model, indicating very weak spontaneous wave emission for this initial condition. 
This is in agreement with the findings of \citet{Chouksey:22}, who showed that spontaneous wave emission is generally weak for an initial perturbation similar as the one we use here. When perturbing the jet with a so-called ageostrophic unstable mode, spontaneous wave emission can become strong at large Rossby numbers. 
However, for simplicity, we only consider the case of weak wave emission in this study.

The OBTA imbalance follows the same pattern as in the shallow water model.
For large Rossby numbers, the OB and OBTA imbalance are the same, while for smaller Rossby numbers, the OBTA imbalance reaches a plateau and does not get smaller than a certain value.
This plateau is again given by the OBTA balancing error of the initial condition (not shown here) and can be decreased by increasing the number of time-averaging chunks.
This implies that the results from Sec.~\ref{sec:geo_proj_error} and Sec.~\ref{sec:res_obta_vs_ob} also hold for the non-hydrostatic model, 
i.e. the geostrophic projection error $\delta_\text{proj}$ and the OBTA balancing error $\epsilon_\text{dev}$ converge to zero when increasing the number of equidistant time chunks.

\section{Summary and conclusions}
\label{sec:conclusions}

To quantify and to understand processes such as spontaneous wave emission under realistic conditions, 
an accurate flow decomposition method 
is needed which can be applied to 
ocean models that that include irregular lateral and vertical boundaries as well as a varying Coriolis parameter.
Until now, it has not been possible to apply existing balancing methods such as OB or NNMD to such realistic models, because 
of the need of Fourier transforms for geostrophic projections.
With our  modification of OB presented here, where the projection  is achieved through a time-averaging procedure (OBTA), it becomes possible to apply OB to realistic models.

In this paper, we demonstrated that OBTA converges to the original OB method. 
For this demonstration, we used idealized setups of the shallow water model and an non-hydrostatic model without lateral boundaries and with a constant Coriolis parameter using different initial
conditions and a wide range of Rossby numbers $Ro$.
For different initial conditions and across all $Ro$, the OBTA imbalance consistently converge towards the OB imbalance as the number of iterations is increased or the time-averaging error decreased. We also applied 
for the first time both OB and OBTA to a non-hydrostatic model with similar results as in the shallow water model.

The time-averaging error in OBTA can be reduced by increasing the time-averaging period. However, this introduces a major drawback of OBTA, as it drastically increases the computational cost of the already computationally expensive OB method. 
However, we find that
the computational cost of OBTA can be significantly reduced by performing multiple shorter time averages instead of a single long one. 
The time-averaging periods should be equidistantly distributed in the range of half an inertial period and a full inertial period, where two to three time chunks seems enough to obtain similar results with OBTA as for OB.

While the OBTA method can  be applied now to more realistic models, the effects of these features on OB may present interesting
challenges.
The assumption that the slow component of the linear system is adiabatically transformed into the slow component of the nonlinear system during the forward ramping may no longer hold. 
This could for example be the case if the slow component interacts with the boundaries and generates fast waves. 
If these fast waves are eliminated during the backward ramping and generated during the forward ramping, the final balanced state may not be the best balanced state, even though the OB iteration converges. 
Interactions of the slow mode with boundaries
and possible generation of fast waves, including Kelvin waves are the scope of a future study.
 

For the case of a varying Coriolis parameter, the geostrophic mode, which is constant in time in the linear system with constant Coriolis parameter, becomes the Rossby mode, which slowly varies in time. Hence, the time-averaging procedure will no longer converge. However, since the frequency of Rossby waves is much smaller than the frequency of inertia-gravity waves, the inertia-gravity waves are much more strongly damped than the Rossby waves during the time-averaging. Therefore, the projection with time-averaging might still be a good approximation of the projection onto the Rossby mode. In case the damping of the Rossby mode is too strong, an alternative method needs to be found to project onto the Rossby mode. In the meantime, optimal balance with time-averaging comes forth as a new flow decomposition tool that can be applied to a wide range of complex flows without compromising the precision of the decomposed fields.



\backsection[Funding]{This paper is a contribution to subprojects L2 and W6 of the Collaborative Research Centre 
TRR 181 ``Energy Transfers in Atmosphere and Ocean'' funded by the Deutsche Forschungsgemeinschaft 
(DFG, German Research Foundation) under project number 274762653. }

\backsection[Declaration of interests]{The authors report no conflict of interest.}

\backsection[Data availability statement]{
The source code of all experiments conducted in this paper are publicly available in a GitHub repository. The GitHub repository can be accessed at the following URL:
\url{https://github.com/Gordi42/Flow-Decomposition-With-OBTA}}

\backsection[Author ORCIDs]{
S. Rosenau, https://orcid.org/0000-0002-8157-2574;
M. Chouksey, https://orcid.org/0000-0001-7161-9116;
C. Eden, https://orcid.org/0000-0002-0925-2170}


\appendix
\section{Estimation of the OBTA error}
\label{appendix:OBTA_error}

We can express the time-averaging projector ${\mathsfbi P}_T$ by using the time-averaging error $\delta$ as
\begin{equation}
  {\mathsfbi P}_T = {\mathsfbi P}_0 + \delta {\mathsfbi P}_\text{IG}
  \hspace{0.5cm}
  \overset{{\mathsfbi P}_0 + {\mathsfbi P}_\text{IG} = 1}{\implies} \hspace{0.5cm}
  1 - {\mathsfbi P}_T = (1 - \delta) {\mathsfbi P}_\text{IG}
\end{equation}
with the inertial-gravity wave mode projector ${\mathsfbi P}_\text{IG}: \mathbb C^3 \longrightarrow \mathbb E_\text{IG} = \mathbb E_1 \times \mathbb E_2$
\begin{equation}
  {\mathsfbi P}_\text{IG} = {\mathsfbi P}_1 + {\mathsfbi P}_2
\end{equation}
For the calculation of the OBTA error we assume that the order of magnitude of the residual does not change when applying the ramping operators $\mathcal R^f$, and $\mathcal R^b$
\begin{eqnarray}
  {\mathsfbi P}_\text{IG} \bcdot \mathcal R^b (\boldsymbol{z}_0 + \boldsymbol{z}_s + \boldsymbol{z}_r) &=& \textit{O}(\boldsymbol{z}_r) \\
  \mathcal{R}^f ( \boldsymbol{z}_0 + \boldsymbol{z}_r ) &=& \mathcal R ^f (\boldsymbol{z}_0) + \textit{O} (\boldsymbol{z}_r)
\end{eqnarray}
Using these assumptions, we can express the states after each OBTA iterations (denoted with tilde) in terms of the states after the corresponding OB iteration
\begin{eqnarray}
  \boldsymbol{\Tilde z}^\text{base} &=&
  \boldsymbol{z}^\text{base} + \delta (\boldsymbol{z}_s + \boldsymbol{z}_r)
  \\
  \boldsymbol{\Tilde z}^{(ii)} &=&
  \mathcal{R}^b ( \boldsymbol{z}_0 + \boldsymbol{z}_s + \boldsymbol{z}_r) = \boldsymbol{z}^{(ii)}
  \\
  \boldsymbol{\Tilde z}^{(iii)} &=&
  {\mathsfbi P}_T \bcdot \boldsymbol{\Tilde z}^{(ii)} =
  {\mathsfbi P}_0 \bcdot \boldsymbol{z}^{(ii)} +
  \delta {\mathsfbi P}_\text{IG} \bcdot \mathcal R^b (\boldsymbol{z}_0 + \boldsymbol{z}_s + \boldsymbol{z}_r) =
  \boldsymbol{z}^{(iii)} + \delta \textit{O}(\boldsymbol{z}_r)
  \\
  \boldsymbol{\Tilde z}^{(iv)} &=& 
  \mathcal R^f (\boldsymbol{\Tilde{z}}^{(iii)}) =
  \mathcal R^f (\boldsymbol{{z}}^{(iii)} + \delta \textit O (\boldsymbol{z}_r)) = \boldsymbol{{z}}^{(iv)} + \delta \textit O (\boldsymbol{z}_r)
  \\
  \boldsymbol{\Tilde z}^{(v)} &=&
  (1 - {\mathsfbi P}_T) \boldsymbol{\Tilde{z}}^{(iv)} + \boldsymbol{\Tilde{z}}^\text{base} = 
  (1 - \delta) {\mathsfbi P}_\text{IG} \bcdot \boldsymbol{\Tilde{z}}^{(iv)} + \boldsymbol{\Tilde{z}}^\text{base} \\
  &=& \boldsymbol{z}^{(v)} + \delta \left( 
  \boldsymbol{z}_r + \boldsymbol{z}_s - {\mathsfbi P}_\text{IG} \bcdot \boldsymbol{z}^{(iv)} + (1 - \delta) {\mathsfbi P}_\text{IG} \bcdot \textit O (\boldsymbol{z}_r)
  \right)
  \label{eq:error_of_obta_very_long_form}
\end{eqnarray}
with
\begin{equation}
  \boldsymbol{z}^{(v)} = {\mathsfbi P}_\text{IG} \bcdot \boldsymbol{z}^{(iv)} + \boldsymbol{z}^\text{base}
  \label{eq:ob5_alternative_version}
\end{equation}
Note that Eq.~(\ref{eq:ob5_alternative_version}) is equivalent to the definition in Eq.~(\ref{eq:ob5_with_geostrophic}). To further simplify Eq.~(\ref{eq:error_of_obta_very_long_form}), we define the OB error as the deviation between the true slaved mode and the obtained slaved mode after one OB iteration
\begin{equation}
  \boldsymbol{\epsilon}_\text{OB} =
  \boldsymbol{z}^{(v)} - \boldsymbol{z}_0 - \boldsymbol{z}_s
  \hspace{0.75cm}
  \overset{\text{(Eq. \ref{eq:ob5_alternative_version})}}{\implies}
  \hspace{0.75cm}
  {\mathsfbi P}_\text{IG} \bcdot \boldsymbol{z}^{(iv)} = 
  \boldsymbol{z}_s + \boldsymbol{\epsilon}_\text{OB}
  \label{eq:OB_error}
\end{equation}
Inserting Eq.~(\ref{eq:OB_error}) into Eq.~(\ref{eq:error_of_obta_very_long_form}) eliminates the error dependency on the slaved mode
\begin{equation}
  \boldsymbol{\Tilde z}^{(v)} = 
  \boldsymbol{z}^{(v)} + \delta \left( 
    \boldsymbol{z}_r - \boldsymbol{\epsilon}_\text{OB} + (1 - \delta) {\mathsfbi P}_\text{IG} \bcdot \textit{O} (\boldsymbol{z}_r)
  \right)
\end{equation}
The term $\delta \boldsymbol{\epsilon}_\text{OB}$ is always smaller than the OB error itself and can thus be safely neglected. All other terms are of order $\textit{O}(\boldsymbol{z}_r)$ or smaller. Hence, the OBTA state after the fifth step is given by
\begin{equation}
  \boldsymbol{\Tilde z}^{(v)} = 
  \boldsymbol{z}^{(v)} + \delta 
  \textit{O} (\boldsymbol{z}_r)
\end{equation}
Note that all terms that contribute to the OBTA error in Eq.~(\ref{eq:error_of_obta_very_long_form}) are part of the inertia-gravity wave space $\mathbb E_\text{IG}$. 
Hence the boundary condition that the geostrophic mode of the initial state and the balanced state are the same is always fullfilled.


\bibliographystyle{jfm}

\bibliography{references}  


\end{document}